%% file: main.tex
\documentclass[twocolumn,nofootinbib,table]{aastex631}

\usepackage{natbib}
\usepackage[utf8]{inputenc}
\usepackage{amsmath}
\usepackage{amsmath,units}
\usepackage{amsmath,bm}
\usepackage{booktabs}
\usepackage{makecell} 
\usepackage{xcolor}
\usepackage{graphicx}
\usepackage{acronym}   
\usepackage{xspace}
\usepackage{enumitem} 
\usepackage{calc}
\usepackage[caption=false]{subfig}
\usepackage{import}
\usepackage{mathtools} 

\newcommand\COMPAS{{\sc{COMPAS}}\xspace}
\newcommand*\diff{\mathop{}\!\mathrm{d}}
\newcommand{\MSFR}{\ensuremath{{M}_{\rm{SFR}}}\xspace}
\newcommand{\ts}{\ensuremath{{t}_{\rm{s}}}\xspace}
\newcommand{\tsup}{\textsuperscript}
\newcommand{\Vc}{\ensuremath{{V}_{\rm{c}}}\xspace}

\newcommand{\monei}{\ensuremath{m_{1,\rm{i}}}\xspace}
\newcommand{\mtwoi}{\ensuremath{m_{2,\rm{i}}}\xspace}

\newcommand{\ai}{\ensuremath{a_{\rm{i}}}\xspace}
\newcommand{\qi}{\ensuremath{q_{\rm{i}}}\xspace}
\newcommand{\Zi}{\ensuremath{Z_{\rm{i}}}\xspace}
\newcommand{\vk}{\ensuremath{v_{\rm{k}}}\xspace}
\newcommand{\thetak}{\ensuremath{{\theta}_{\rm{k}}}\xspace}

\newcommand{\ei}{\ensuremath{{e}_{\rm{i}}}\xspace}

\newcommand{\Msun}{\ensuremath{\,\rm{M}_{\odot}}\xspace}
\newcommand{\Zsun}{\ensuremath{\,\rm{Z}_{\odot}}\xspace}
\newcommand{\kms}{\ensuremath{\,\rm{km}\,\rm{s}^{-1}}\xspace}
\newcommand{\AU}{\ensuremath{\,\mathrm{AU}}\xspace}

\usepackage{array}
\newcolumntype{E}[1]{>{\raggedright\let\newline\\\arraybackslash\hspace{0pt}}m{#1}}
\newcolumntype{F}[1]{>{\centering\let\newline\\\arraybackslash\hspace{0pt}}m{#1}}
\newcolumntype{G}[1]{>{\raggedleft\let\newline\\\arraybackslash\hspace{0pt}}m{#1}}

\acrodef{ANN}{artificial neural network}
\acrodef{DCO}{double compact object}
\acrodef{MCMC}{Markov Chain Monte Carlo}
\acrodef{MSSFR}{metallicity-specific star formation rate}

\definecolor{orange-red}{rgb}{1.0, 0.27, 0.0}
\newcommand{\resp}[1]{#1}

\begin{document}

\title{\resp{Surrogate forward models for population inference on compact binary mergers}}

\author{Jeff Riley}
\affiliation{School of Physics and Astronomy, Monash University, Clayton, Victoria 3800, Australia}
\affiliation{ARC Centre of Excellence for Gravitational Wave Discovery -- OzGrav, Australia}
\author{Ilya Mandel}
\affiliation{School of Physics and Astronomy, Monash University, Clayton, Victoria 3800, Australia}
\affiliation{ARC Centre of Excellence for Gravitational Wave Discovery -- OzGrav, Australia}

\correspondingauthor{Jeff Riley}
\email{jeff.riley@monash.edu}

\begin{abstract}
Rapidly growing catalogs of compact binary mergers from advanced gravitational-wave detectors allow us to explore the astrophysics of massive stellar binaries.  Merger observations can constrain the uncertain parameters that describe the underlying processes in the evolution of stars and binary systems in population models.  In this paper, we demonstrate that binary black hole populations -- namely, detection rates, chirp masses, and redshifts -- can be used to measure cosmological parameters describing the redshift-dependent star formation rate and metallicity distribution.   We present a method that uses artificial neural networks to emulate binary population synthesis computer models, and construct a fast, flexible, parallelisable surrogate model that we use for inference. 
\end{abstract}


\import{}{intro.tex}

\import{}{method.tex}

\import{}{results.tex}

\import{}{conclusion.tex}

\import{}{ack.tex}

\section*{Data availability}
The data underlying this article is available from the corresponding author upon request.

\bibliographystyle{aasjournal}
\bibliography{bib.bib}


\clearpage
\appendix
\import{}{appendix-ANNs.tex}

\clearpage
\import{}{appendix-COMPAS_fiducial.tex}

\end{document}

%% file: intro.tex
\section{Introduction}\label{sec:intro}

Rapid binary population synthesis software packages (e.g., BSE \citep{Hurley_2002}, StarTrack \citep{Belczynski_2008}, binary\_c \citep{Izzard_2018}, SEVN \citep{Iorio_2022}, \COMPAS \citep{COMPAS_2022}, cosmic \citep{Breivik_2020}) have a number of free parameters that allow modellers to:

\begin{itemize}
\setlength{\itemindent}{6mm}
  \item set the initial conditions of the simulation (e.g. initial stellar mass, metallicity, separation (for binary stars), etc.), and
  \item determine the nature of some of the (simulated) physical processes that a star, single or binary, undergoes during its evolution through time (e.g. wind mass loss rate, supernova remnant masses and kicks, mass transfer and common envelope efficiency for binary stars).
\end{itemize}

Many of these free parameters and the physical states and processes they represent are so far not well constrained by observation.

Our ultimate goal is to develop a method to determine constraints for (some of) the physical states and processes modelled by various modelling and simulation software packages, thus not only learning more about the physical nature of these states and processes, but also improving our ability to model them. A na\"ive, brute-force, method to determine such constraints is to generate a synthetic state space by integrating over all possible values of initial conditions and evolutionary parameters, then search that synthetic state space for states that correspond to observed data. However, even an indicative selection of that parameter space is very large (e.g., \citet{Broekgaarden_2022} considered 560 distinct model variations), and generating such a large state space with current modelling tools is computationally intractable.

As a first step towards achieving our goal, we develop a tool that can quickly simulate a state space large enough to allow us to infer astrophysical constraints. We can do that by generating a smaller number of states using a targeted selection of initial conditions and evolutionary parameters, and using that simulated state space as a set of training examples to teach an interpolant how to map initial conditions and evolutionary parameters to the final state, thus avoiding the computational time of calculating the intermediate steps. Once we have a working tool we can then generate a very large simulated state space and search that state space for states that match observed data. Here, as proof of the concept, we describe the construction of the tool for a reduced space of parameters: four parameters governing the cosmic \ac{MSSFR} in the model of \citet{Neijssel_2019}.

The remainder of this paper is organised as follows. Section~\ref{sec:method} presents a description of the tool we construct for this proof-of-concept study, and the method used to train the tool.  We present and discuss our results in Section~\ref{sec:results}.  We provide some concluding remarks in Section~\ref{sec:conclusion}.

%% file: method.tex
\section{Method}\label{sec:method}

\subsection{Overview}\label{sec:method_overview}

The \COMPAS population synthesis software suite \citep{Stevenson_2017,Vigna-Gomez_2018,COMPAS_2022} allows users to generate synthetic populations of binary systems. The \COMPAS suite includes post-processing tools that allow users to, amongst other things, calculate merger rates, detection rates of mergers, etc., for \acp{DCO} that will merge within the age of the universe \citep{Neijssel_2019,Barrett_2018,COMPAS_SOFTWARE_2022}.

For this proof-of-concept study we develop a surrogate model for \COMPAS that provides fast predictions of detection rates of merging \acp{DCO} as a function of chirp mass and merger redshift. Surrogate models have been used to emulate various aspects of the simulated evolution of binary systems and to infer constraints on simulation parameters, including the evolution of individual binary systems (e.g. \citet{Lin_2021}, using a Gaussian Process emulator \citep{Rasmussen_Williams_2005} and classifier) and populations of binaries (e.g. \citet{Barrett_2016}, \citet{Taylor_Gerosa_2018}, \resp{\citet{Wong_2019}, \citet{Cheung_2022} using Gaussian Process emulators or normalising flows, and \citet{Wong_2021} using deep-learning enhanced hierarchical Bayesian analysis}).

We chose detection rates because these are calculated by the \COMPAS post-processing tools after the evolution of the population of binary systems by convolving binary evolution outcomes with the \ac{MSSFR}.  This makes generating example (training) data for our surrogate model by varying \ac{MSSFR} parameters much less onerous than it would be if we needed to evolve new populations of binaries whenever our astrophysical parameters are varied.

We construct an interpolant that, given a set of input values, predicts a detection rate matrix for the merger of \acp{DCO}. Input to the interpolant is a set of values for four \ac{MSSFR} parameters chosen for this study (see Section~\ref{sec:method_parameters}). The interpolant outputs detection rates binned into a matrix of 113 chirp mass bins by 15 redshift bins (see Section~\ref{sec:method_binned_rates}). The detection rates generated by the interpolant are combined rate predictions for all \acp{DCO}: binary black holes, binary neutron stars, and mixed neutron star -- black hole binaries.  The interpolant we construct is comprised of a matrix of simple feed-forward \acp{ANN}, with each \ac{ANN} trained to predict the detection rate for a single cell of the resultant detection rate matrix.

We first simulate the evolution of 512 million binary systems using \COMPAS, resulting in a population of approximately 1.3 million \acp{DCO} that merge within the age of the Universe, $\sim 13.8$ Gyr.

The \COMPAS modelling tool has many configurable options that allow the user to change initial conditions and evolutionary parameters. The simulation for this study was performed using the \COMPAS fiducial values for all configurable options \resp{(see Appendix~\ref{sec:appendix_COMPAS_fiducial}, Table~\ref{tab:app_COMPAS_fiducial}}; described in detail in \citet{COMPAS_2022}), and for our purposes we consider the population of binary systems simulated by \COMPAS to be indicative of the population of binary systems in the Universe.

The \COMPAS suite includes tools to take a population of binary systems produced by \COMPAS, integrate over the known initial condition distribution and the cosmological star formation history, and produce a population of merging \acp{DCO} with individual masses and merger redshifts, to which observational selection effects can be applied, converting the population into a prediction for the observable merger rate over the population of \acp{DCO}.

We construct a grid of six different values for each of the four \ac{MSSFR} parameters under study (see Section~\ref{sec:method_parameters}), and for each entry in the grid we us the \COMPAS tools described above to calculate a detection rate matrix for the population of merging \acp{DCO}, resulting in 1,296 detection rate matrices, which we use to create the training data for the \acp{ANN} that comprise our interpolant. The calculated detection rate matrices contain, for each bin in the chirp mass -- redshift space, the expected detection rate (detections per year) for an interferometer network with an indicative sensitivity of the LIGO O3 run \citep{scenarios}.

For each cell in the detection rate matrix we construct and train an \ac{ANN} that predicts the detection rate for that cell, given values for the four \ac{MSSFR} parameters -- this matrix of \acp{ANN} comprises our interpolant\footnote{\label{constant_cells}Some cells in the detection matrix may not require the use of an \ac{ANN} -- see Section~\ref{sec:method_interpolant}.}. We chose a matrix of simple \acp{ANN} rather than a single, large \ac{ANN} for several reasons:

\begin{enumerate}[label=(\roman*)]
  \item Training a large \ac{ANN} to learn $>1500$ relationships is difficult and time-consuming, whereas training a small \ac{ANN} to learn a single relationship is easier and faster.
  \item A large \ac{ANN} is likely to be less accurate because the network will try to get most relationships close, whereas the single, small \ac{ANN} just needs to get one relationship right.
  \item Over-fitting in a large \ac{ANN} learning multiple relationships is more likely than a single, small \ac{ANN} learning a single relationship. It is likely that some of the $>1500$ different relationships to be learned will be easier to learn than others.  Training a large \ac{ANN} to learn multiple relationships could result in the relationships that are easier to learn being over-fitted because the \ac{ANN} needs to be trained for longer to learn the more difficult relationships.
  \item \acp{ANN} for different cells can have different architectures - some relationships might require more, or fewer, nodes and/or layers.
  \item While other global surrogate models could provide computational cost savings by utilising the regularity (smoothness) of the output over the input parameter space, assumptions of regularity are risky when some astrophysical parameters could lead to bifurcations in the outputs.
  \item We can replace the \ac{ANN} for some cells with a different model without affecting all cells -- some relationships may not need to be modelled by an \ac{ANN}: a simple function fit (possibly even a constant value) might be sufficient.
  \item We can retrain the \acp{ANN} for individual cells as necessary, without the need to retrain all cells.
  \item We can improve performance by running individual \acp{ANN} in parallel on different CPUs.
\end{enumerate}

\subsection{MSSFR parameters}\label{sec:method_parameters}

The parameters we chose to vary for this study are four free parameters for the calculation of the \ac{MSSFR} in the phenomenological model of \citet{Neijssel_2019}. In this model, the \ac{MSSFR} is split into two parts, the star formation rate (SFR) and the metallicity distribution:

\begin{equation}
  \frac{\diff^3 \MSFR}{\diff \ts \diff \Vc \diff Z}(z) =
  \frac{\diff^2 \MSFR}{\diff \ts \diff \Vc }(z)
  \times 
  \frac{\diff P }{\diff Z}(z),
\end{equation}

\noindent
where the SFR is given by:

\begin{equation}
  \frac{\diff^2 \MSFR}{\diff \ts \diff \Vc} =
  a\frac{(1 + z)^b}{1 + (\frac{(1 + z)}{c})^d}
  \mathrm{M}_\odot\ \mathrm{yr}^{-1}\ \mathrm{Mpc}^{-3},
\end{equation}

\bigskip\noindent
and we use a log-normal distribution in metallicity at each redshift (cf.~the skewed log-normal model proposed by \citet{vanSon_2022}):

\begin{equation}
  \frac{\diff P }{\diff Z}(z) =
  \frac{1}{Z\sigma\sqrt{2\pi}}e^{-\frac{(\ln(Z)-\ln(\langle Z \rangle) +\sigma^2/2)^2}{2\sigma^2}},
\end{equation}

\bigskip\noindent
with a redshift-independent standard deviation $\sigma$ in $\ln(Z)$ space around a redshift-dependent mean $\mu$ of $\ln(Z)$ given by:

\begin{equation}
  \langle Z \rangle =
  e^{(\mu + \frac{\sigma^2}{2})},
\end{equation}

\bigskip\noindent
with mean metallicity parametrised as in \citet{Langer_2006}:

\begin{equation}
  \langle Z(z) \rangle =
  Z_{0}10^{\alpha{z}}
\end{equation}

\bigskip\noindent
We vary SFR parameters $a$ and $d$, and the metallicity distribution parameters $\alpha$ and $\sigma$, in this study, while fixing $b=2.77$, $c=2.9$, and $Z_0=0.035$.  \resp{\citet{Neijssel_2019} demonstrated that a range of star-formation history models can be mimicked by varying only this subset of \ac{MSSFR} parameters.  We therefore choose to vary the same parameters for consistency with their analysis.  Moreover, reducing the number of parameters allows us to limit computational complexity for this proof-of-principle study.}

\subsection{Binned rates}\label{sec:method_binned_rates}

As noted in Section~\ref{sec:method_overview}, the detection rates generated by our interpolant are binned into a matrix of 113 chirp mass bins by 15 redshift bins.

The redshift bins are constructed as fixed-width bins, with the lower edge of the first bin at redshift 0.0 and each bin having a width of 0.1 -- the upper edge of the final (15\tsup{th}) bin is at redshift 1.5.

The chirp mass bins are constructed as variable-width bins, with the lower edge of the first bin at $L_1=0.0 \Msun$ and the upper edge at $L_2 = 0.5\Msun$. The remaining chirp mass bin edges fall at

\begin{equation}
  L_{i+1}=\frac{41}{39} L_i, \ \ 2 \leq i \leq 112.
\end{equation}

Thus, the width of the 2\tsup{nd} to 112\tsup{th} bins is equal to 5\% of the median chirp mass for the bin. The final (113\tsup{th}) bin extends to infinity.

\subsection{The Interpolant}\label{sec:method_interpolant}

\begin{figure}
  \centering
  \includegraphics[width = 6.0cm]
    {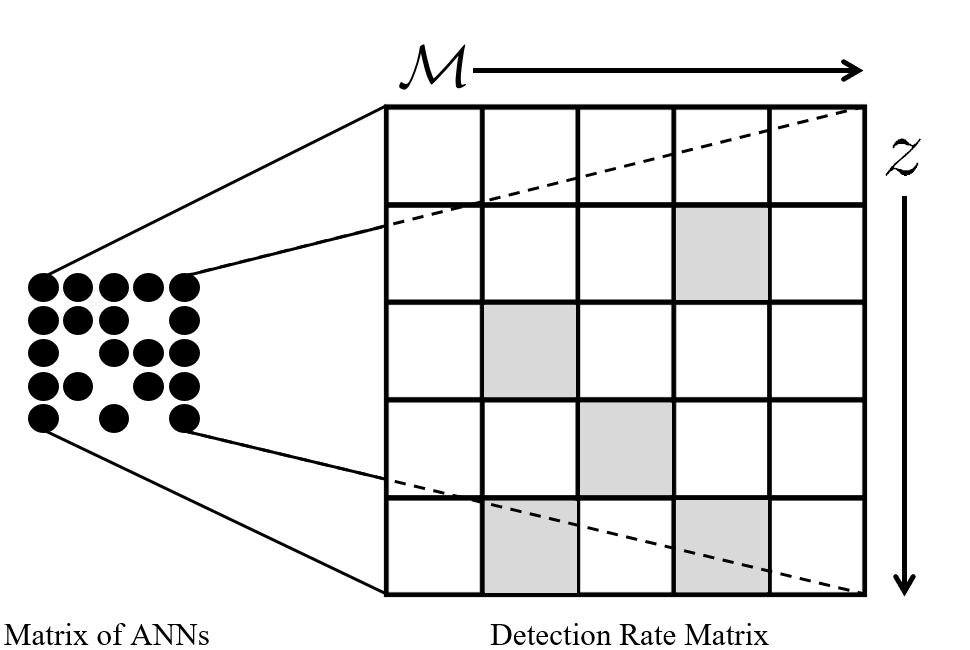}
    \caption{The Interpolant}
    \label{fig:method_interpolant}
\end{figure}

As noted in Section~\ref{sec:method_overview}, the interpolant we construct is comprised of a matrix of \acp{ANN}. For this study, all \acp{ANN} in the matrix have identical architecture, though that is not a requirement of the method: each \ac{ANN} is independent of the others, and the architecture of individual \acp{ANN} could be varied. Nominally, there will be a corresponding \ac{ANN} for each cell in the detection rate matrix, but there may be some cells for which an \ac{ANN} is not required. For example, if a given cell in the detection rate matrix is known to be a constant value (e.g.~0), training an \ac{ANN} is both not useful and not necessary - we can instead replace the \ac{ANN} for that cell with (e.g.) a simple equation and skip training an \ac{ANN}. A schematic diagram of the interpolant is shown in Figure~\ref{fig:method_interpolant}. In Figure~\ref{fig:method_interpolant}, \acp{ANN} are trained to predict values only for the unshaded cells in the detection rate matrix: values for the shaded cells will be predicted using some other model, as discussed above.

\subsubsection{The Artificial Neural Networks}\label{sec:method_interpolant_ANNs}

\begin{figure}
  \centering
  \includegraphics[width = 5.0cm]
    {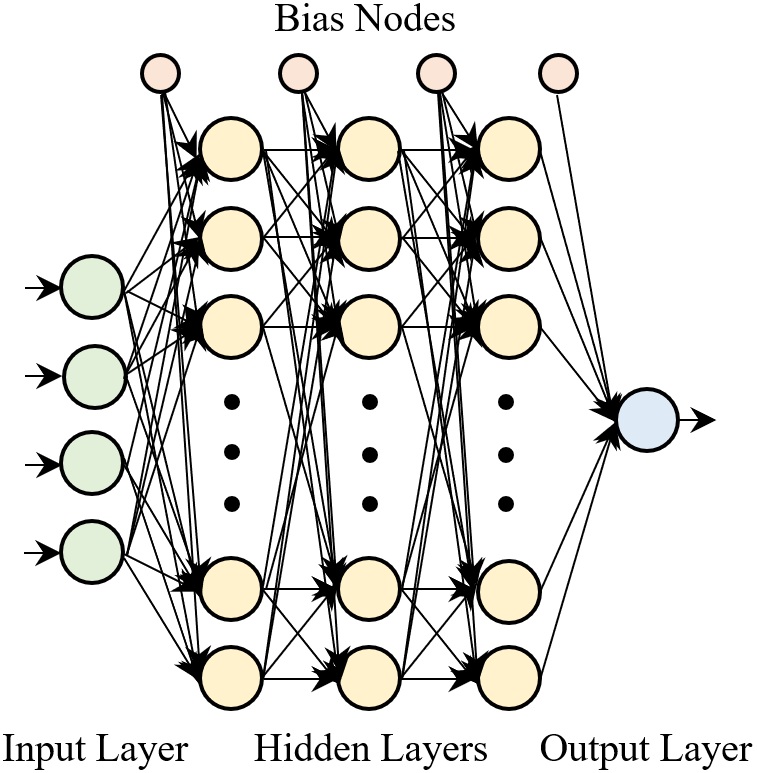}
    \caption{ANN architecture}
    \label{fig:method_ANN_architecture}
\end{figure}

Each \ac{ANN} is a fully-connected feed-forward network of artificial neurons (nodes), with an input layer, three hidden layers, and an output layer. The input layer consists of four input nodes (one for each \ac{MSSFR} parameter), the output layer consisting of a single node (to indicate the predicted detection rate for the corresponding matrix element), and each of the three hidden layers contains 64 nodes.  The hidden and output layers have an associated bias node that provides each node in the respective layer with a bias weight. A schematic diagram of a representative \ac{ANN} is shown in  Figure~\ref{fig:method_ANN_architecture}.

While there are some generally accepted rules of thumb, the selection of \ac{ANN} architecture is very often arbitrary. In this case the architecture was chosen to produce relatively robust results quickly – no attempt was made to optimise the architecture for the best performance, either with respect to accuracy or speed.  The input layer and output layer are fixed (the four \ac{MSSFR} parameters are the inputs, and since the \ac{ANN} is a regressor, not a classifier, we have a single, continuous, output node), and the choice of three hidden layers, each with 64 nodes, was a brute-force rather than finessed approach. Per the Universal Approximation Theorem\footnote{https://en.wikipedia.org/wiki/Universal\_approximation\_theorem}, a shallow neural network (with a single, wide, hidden layer) can approximate any function – after a little experimentation with different architectures the architecture described above was found to produce good enough results reasonably quickly. While there is no universally accepted rule as to how many layers constitute a deep \ac{ANN} (how many grains make a heap?\footnote{https://en.wikipedia.org/wiki/Sorites\_paradox}), the architecture implemented here probably qualifies as a deep network (because the networks have more than two hidden layers) - but it is essentially a simple feed-forward \ac{ANN} (albeit, given the number of nodes on each layer, a large simple feed-forward \ac{ANN}).

\subsubsection{Training}\label{sec:method_interpolant_training}

Training an \ac{ANN} requires many examples of the relationship(s) to be learned (e.g. example outputs for given inputs) - anywhere from hundreds to millions depending upon the complexity of the relationship(s) to be learned. By designing the interpolant as a matrix of independent, small \acp{ANN}, we have minimised the number, and complexity, of the relationships to be learned by a single \ac{ANN}, and as a result the size of the training data set required is significantly reduced.

The data used to train the \acp{ANN} was constructed by using the \COMPAS post-processing tools to calculate detection rate matrices for 1,296 combinations of the \ac{MSSFR} parameters under study - 6 different values for each of the 4 parameters, collectively labeled $\lambda$, as shown in Table~\ref{table:method_lambda_training_values}. The trained interpolant is expected to be able to interpolate between the bounds of each of the constituent parameters of $\lambda$.

\begin{table}[ht!]
    \centering
    \caption{$\lambda$ training values}
    \begin{tabular}{|F{1.0cm}|F{6.5cm}|}
        \hline
        \bm{$\lambda_x$} & \textbf{Values} \\
        \hline
        $\alpha$ & [-0.500,~-0.400,~-0.300,~-0.200,~-0.100,~-0.001] \\
        \hline
        $\sigma$ & [~0.100,~~0.200,~~0.300,~~0.400,~~0.500,~~0.600] \\
        \hline
        $a_{SF}$ & [~0.005,~~0.007,~~0.009,~~0.011,~~0.013,~~0.015] \\
        \hline
        $d_{SF}$ & [~4.200,~~4.400,~~4.600,~~4.800,~~5.000,~~5.200] \\
        \hline
    \end{tabular}
    \label{table:method_lambda_training_values}
\end{table}

We chose the ranges of values based on the preferred values from \citet{Neijssel_2019}, which are the default \COMPAS values for each of the parameters - the ranges are a spread around the preferred values.

The question of how many training examples are needed to train an \ac{ANN} effectively has no simple answer. We need a sample of data that is representative of the problem we are trying to solve (in our case, constructing an interpolant for \ac{DCO} merger detection rates) and, in general, the examples in the sample should be independent and identically distributed. We are training \acp{ANN} which we hope will learn to correctly map input data to output data, and moreover will learn to interpolate and map input data \resp{that} they have not seen during training - that is, we hope our \acp{ANN} will learn the relationships between the input data and output data. We need sufficient training data to reasonably capture the relationships that (we expect) exist both between input features, and between input features and output features.

The dataset described above contains a single example of the 1,296 mappings (relationships) that we will present to the \acp{ANN}. We know that a single example is almost certainly not sufficient to correctly train the \acp{ANN} - with too few training examples we risk over-fitting the \acp{ANN}, and possibly compromising their ability to generalise beyond the training dataset (but also see Appendix~\ref{sec:appendix_ANN_training_performance}). In fact, we know that the training data are not perfectly accurate because the \COMPAS binary population represents a Monte Carlo sampler over initial conditions.  Therefore, training data suffer from statistical sampling uncertainty. To account for this, we create two new training datasets which we will use to train and test the networks.  The results will give us an indication of the number of training examples required. For each new training dataset, we use bootstrap sampling of \COMPAS outputs to create a number of new detection rate matrices. 

The first sampled dataset contains 10 different bootstrapped examples for each of the 1,296 mappings, resulting in a training dateset with 12,960 matrices, and the second contains 100 different bootstrapped examples for each of the 1,296 mappings, resulting in a training dataset of 129,600 matrices.

As discussed previously, we don't need to train \acp{ANN} for cells in the detection rate matrix that we know to contain constant rates. For each training dataset we inspect all training examples and determine which cells in the matrix are constant across all examples - \acp{ANN} are not required for those cells. We found that for the 1,695 cells in the detection rate matrix we only needed to train 292 \acp{ANN} for the smaller training dataset, and 293 \acp{ANN} for the larger training dataset - for all other cells the detection rate across all training examples was below our threshold for zero (0.001 detections per year).

Training of the interpolant was conducted using the Keras API \citep{Chollet_2015} for Python \citep{Van-Rossum_Drake_2009}. Keras is a high-level neural networks API that uses the Tensorflow machine learning platform \citep{Abadi_2015}.

We used a fairly na\"ive training methodology. Just as no attempt was made to optimise the architecture for the best performance, no real attempt was made to optimise the training for the best accuracy of the networks - "good enough" accuracy was all we were looking for in this proof-of-concept study. We divided our collection of models with different choices of parameters $\lambda$ and corresponding bootstrapped examples into a training set (80\% of all choices of $\lambda$), and a validation set (the remaining 20\% choices of $\lambda$). Networks were trained using the set of training examples, and tested against the validation set - examples that the network undergoing training had not seen during training. $k$-fold cross-validation was not performed - some non-exhaustive tests were performed using $k$-fold cross-validation with no significant improvement in accuracy, so in the interest of reduced training time we chose not to implement $k$-fold cross-validation.

Each network was initialised with random connection weights and biases, and trained for a maximum of 4,500 epochs (an epoch is the presentation of the entire training set to the network), after which the network was tested on the validation set. Early stopping was enabled - if no improvement in accuracy (on the training data) was evident after a specified number of epochs (since the last improvement), training was halted.

The network attributes (weights and biases) were adjusted during training using the Keras Adam optimiser (a stochastic gradient descent optimisation algorithm), with a custom metric for measuring network performance (\footnote{https://keras.io/api/optimizers/adam/}).

The metric used for measuring the performance of the network was half the Poisson uncertainty of the detection count associated with the bin being predicted assuming 1 year of observations: where the expected (target) detection count for a given bin is $t$, a prediction, $p$, is considered correct if $(t - \frac{\sqrt{t}}{2}) \leq p \leq (t + \frac{\sqrt{t}}{2})$. We used sufficiently large \COMPAS binary populations for this study to ensure that the bootsrapping uncertainty is typically small relative to $\sqrt{t}$: fewer than 0.01\% of training and validation samples fall outside a range of width $\sqrt{t}$. The majority of these outliers were found in specific bins in chirp mass -- redshift space with low \ac{DCO} counts (see Section \ref{sec:results_training_accuracy}) and correspondingly high bootstrapping uncertainty.  The accuracy requirement could have been loosened for such bins, but given the low outlier count, we were comfortable with using this definition of performance.

If the network did not achieve 100\% accuracy on the validation set, it was trained again with a different set of (random) initial weights and biases. If the network did not achieve 100\% accuracy on the validation set after 10 tries at training, the network that achieved the highest accuracy during training was accepted.

\subsubsection{Performance}\label{sec:method_interpolant_performance}

There are three aspects of surrogate model performance that interest us: speed, accuracy, and the ability to generalise (specifically, in our case, interpolate). If we are to use the interpolant to create large synthetic state spaces that we can explore, we need it to be very fast, at least by comparison with existing methods (e.g. the \COMPAS tools described in Section~\ref{sec:method_overview}), to be accurate, and to generalise from the training examples. 

The accuracy of the interpolant, with respect to generating a correct detection rate matrix for a given set of \ac{MSSFR} parameters, is measured during training (see Section~\ref{sec:method_interpolant_training}). Since the validation data are technically used as part of training -- in determining whether training can be stopped or can continue to another iteration -- we conducted several ad-hoc spot checks of the interpolant against results generated using \COMPAS, and in all cases the interpolant matched (within acceptable error bounds) the \COMPAS results, thus confirming the ability to generalise/interpolate.

Measuring the speed of the interpolant is straight-forward. For this study, since the interpolant consists only of \acp{ANN}, the speed of the interpolant is dependent mainly upon the execution (forward propagation) speed of its constituent \acp{ANN}, and how many \acp{ANN} comprise the interpolant - with a small amount of input/output (IO) overhead to read the input data (i.e., the values of $\lambda$), present it to the network, and write the output data.

All speed-related performance tests were conducted on a Hewlett Packard ProLiant ML350p Gen8 with two Intel E5-2650v2 processors running at 2.6GHz, providing 32 threads (with hyper-threading) and 96GB of RAM - performance figures quoted are for that system.

The total elapsed time to create both training datasets was ${\sim} 142$ hours.

Since in this study the \acp{ANN} that comprise our interpolant have identical architectures, if we ignore the IO overhead, the speed of the interpolant is directly proportional to the execution speed of a single \ac{ANN}. Training of the \acp{ANN} that comprise the interpolant is independent, so can be done in parallel.  We measured the average execution speed of a single \ac{ANN} over 1,000,000 executions to be $2.19 \times 10^{-5}$ seconds per execution. If we consider the interpolant to be the serial execution of its (for this study) 293 constituent \acp{ANN} (292 for the smaller dataset), the execution speed of the interpolant (ignoring IO overhead) is $2.19 \times 10^{-5} \times 293 = 6.42 \times 10^{-3}$ seconds. However, since the execution of the constituent \acp{ANN} is independent, the architecture of the interpolant lends itself to parallel execution (it is, in fact, embarrassingly parallel), so the speed of execution of the interpolant could be reduced significantly by spreading the execution of the constituent \acp{ANN} across multiple CPUs/cores.

The total elapsed time to train all \acp{ANN} was ${\sim} 1.5$ hours for the smaller training dataset, and ${\sim} 12$ hours for the larger training dataset.

As noted earlier, our interpolant is only useful if it is significantly faster than existing methods. We measured the average execution speed of the \COMPAS toolset to create a merger detection rate matrix from a population of \acp{DCO} over 50 executions to be 111.97 seconds. The \COMPAS post-processing toolset (to create the detection rate matrix) is not parallelisable.

Our interpolant, if run in serial fashion, is ${\sim} 17,500$ times faster than the \COMPAS toolset, and, na\"ively, could be up to ${\sim} 5$ million times faster if run in a fully parallel fashion (though there would be some overhead in running in parallel). We should note that the \COMPAS post-processing tools are written in Python (but it is unlikely that any optimisation would result in improvements of more than tens of percent).

\subsection{Searching the state space}\label{sec:method_search}

With the interpolant constructed, we shift our focus to showing that the interpolant enables us to infer constraints on our chosen \ac{MSSFR} parameters. To do this we search the state space over the range of values of the parameters we used to train our interpolant -- the interpolant is not guaranteed to be valid outside those ranges -- and seek out models that match the observed data.

First, to validate the method and test the interpolant, we create a mock data set that, for this test, we consider to be representative of the Universe. We use \COMPAS as our forward model to create a population of \acp{DCO} representative of the Universe, and, as noted earlier, we use the \COMPAS fiducial values for all configurable options.

We use the \COMPAS post-processing tools to create mock ``true'' data, $\mathcal{D}$, as follows. We specify a detector noise spectrum (equivalent to an approximate LIGO O3 sensitivity) and choose ``true'' values of the \ac{MSSFR} parameters $\lambda$. We use the \COMPAS tools to generate a \ac{DCO} merger detection rate matrix that represents the universe defined by the \COMPAS fiducial model. We multiply the detection rate matrix by a fixed run duration to produce a matrix of expected detections, and then independently sample mock observations from each chirp mass -- redshift in this matrix following Poisson statistics.

We then use our surrogate model to generate a detection rate matrix at various points in the state space (characterised by varying $\lambda$), multiply this by the same run duration, and calculate the likelihood of observing $\mathcal{D}$ given $\lambda$ at those points as described below. This produces a ``likelihood landscape'' that we can search. Finding the maximum likelihood on the landscape tells us the value of $\lambda$ (the interpolant considers is) most likely to produce our ``true'' data. The likelihood surface (or, in Bayesian language, the posterior on the parameters $\lambda$ under the assumption of flat priors over the range considered) contains information about uncertainty in the \ac{MSSFR} parameter inference and any correlations between these parameters.

Results of the validation of the method are presented in Section~\ref{sec:results_method_validation}.

\subsubsection{Likelihood calculation}\label{sec:method_likelihood_calculation}

We define the likelihood $\mathcal{L}$ that we will see the data $\mathcal{D}$ for a particular set of parameters $\lambda$ as

\begin{equation}
  \log{\mathcal{L}\left( \mathcal{D} |\lambda\right)} = \log{\mathcal{L}\left(N_{\mathrm{obs}}|\lambda\right)} +
  \sum_{i=1}^{N_\mathrm{obs}} \log{p(D_i|\lambda)},
  \label{eq:methods_likelihood}
\end{equation}

\bigskip\noindent
where $N_{\mathrm{obs}}$ is the number of observations in the data $\mathcal{D} \equiv \{D_1, D_2, ... D_{N_{\mathrm{obs}}}\}$.

The first term of Equation~\ref{eq:methods_likelihood} is a Poisson likelihood on there being $N_\mathrm{obs}$ detections,

\begin{equation}
  \log{\mathcal{L}\left(N_{\mathrm{obs}}|\lambda\right)} = N_{\mathrm{obs}}\log(\mu(\lambda)) - \mu(\lambda),
  \label{eq:methods_likelihood_term1}
\end{equation}

\bigskip\noindent
where $\mu(\lambda)$ is the expected number of observations and we omit terms that depend on the data only and therefore disappear on normalisation, such as $\log(N_{\mathrm{obs}}!)$ and permutation coefficients.

The second term of Equation~\ref{eq:methods_likelihood} is comprised of a product of the probabilities of individual detections,

\begin{equation}
  p(D_i|\lambda) = p(z=z_i,\mathcal{M}=\mathcal{M}_{i}|\lambda),
  \label{eq:methods_likelihood_term2}
\end{equation}

\bigskip\noindent
where $p(z=z_i,\mathcal{M}=\mathcal{M}_{i}|\lambda)$ is the entry in the probability distribution matrix $p(z,\mathcal{M}|\lambda)$ in the $z$ and $\mathcal{M}$ bin of the observed $i^{th}$ event.  The probability distribution matrix $p(z,\mathcal{M}|\lambda)$ is constructed by dividing the entries in the matrix of expected detections for a given $\lambda$, generated by multiplying the surrogate model detection rate matrix by the run duration, by the sum of the entries in the matrix, $\mu$.

\subsection{Using gravitational wave observations}\label{sec:method_LVK}

The catalogue of LIGO, Virgo, and KAGRA (LVK) gravitational wave observations (\citet{GWTC-2_1_zenodo}, \citet{GWTC-3_zenodo}) provides us with a number of merger detections from a population of \acp{DCO}.  For each LVK event we know, among other things, the chirp mass of the merging binary and the merger redshift, up to a measurement uncertainty. We only include confident detections with a minimum astrophysical probability $p_{astro} \ge 0.95$; although less confident events could be included \citep[e.g.][]{Farr_2015}, they will typically contribute little information due to greater measurement uncertainties, so we omit them for this proof-of-concept study.

Once we validate our method, we can use our surrogate model and the LVK data to infer real physical constraints on the parameters under study, with the important caveat that we have fixed the astrophysical parameters, and any errors in those could bias the inferred \ac{MSSFR} model values.

\subsubsection{LVK likelihood calculation}\label{sec:method_LVK_likelihood_calculation}

 In practice, for LVK events, we may not know the values $z_i, \mathcal{M}_{i}$ and hence the probability $p(z=z_i,\mathcal{M}=\mathcal{M}_{i}|\lambda)$ for the observed $i^{th}$ event in Equation~\ref{eq:methods_likelihood_term2} perfectly, but may only have $K$ samples from a posterior $p(z_i,M_{c,i}|D_i)$.  For LVK data, we substitute the following for Equation~\ref{eq:methods_likelihood_term2}:
 
\begin{equation}
  p(D_i|\lambda) = \frac{1}{K} \sum_{k=1}^K \frac{p(z=z_i^k, M_c=M_{c,i}^k|\lambda)}{\pi(z_i^k,M_{c,i}^k)},
  \label{eq:methods_likelihood_term2_LIGO}
\end{equation}

\bigskip\noindent
where the subscript $k$ refers to the $k^{th}$ posterior sample among $K$ available samples for event $i$ and $\pi(z,M_c)$ is the prior used by LVK data analysts in the original inference on the observed events \citep[see, e.g.][]{Mandel_2019}.

%% file: results.tex
\section{Results and discussion}\label{sec:results}

\subsection{Interpolant training results}\label{sec:results_training_results}

The detection rate matrix generated by our interpolant consists of 1695 cells: 113 chirp mass bins x 15 redshift bins. Of the 1695 cells in the matrix, only ~17\% (292 for the smaller training dataset, and 293 for the larger training dataset) require an \ac{ANN} to predict the rate for that cell: all other cells have zero rates for all training examples. The time to train the interpolant, and the accuracy achieved during training, are shown below.

\subsubsection{Training times}\label{sec:results_training_times}

Details of the training times for the \acp{ANN} trained are presented in Table~\ref{table:results_interpolant_training_stats}.  As noted in Section~\ref{sec:method_interpolant_performance}, the total elapsed time to train all \acp{ANN} was ${\sim} 1.5$ hours for the smaller training dataset, and ${\sim} 12$ hours for the larger training dataset.

\begin{table}[ht!]
    \centering
    \caption{ANN training statistics}
    \begin{tabular}{|E{4.0cm}|G{1.4cm}|G{1.4cm}|}
        \hline
        & \multicolumn{2}{c|}{\textbf{\#examples}} \\
        \cline{2-3}
        \multicolumn{1}{|c|}{\textbf{Training Statistic}} & \multicolumn{1}{c|}{\textbf{10}} & \multicolumn{1}{c|}{\textbf{100}} \\
        \hline
        Lowest train time (min) & 0.26 & 1.37 \\
        \hline
        Highest train time (min) & 42.98 & 301.37 \\
        \hline
        Average train time (min) & 9.85 & 74.1 \\
        \hline
        Average tries to train & 5.76 & 7.18 \\
        \hline
        Fewest tries to train & 1 & 1 \\
        \hline
        Most tries to train & 10 & 10 \\
        \hline
        Number of networks trained & 292 & 293 \\
        \hline
    \end{tabular}
    \label{table:results_interpolant_training_stats}
\end{table}

\subsubsection{Interpolant accuracy}\label{sec:results_training_accuracy}

As discussed in Section~\ref{sec:method_interpolant_training}, each of the constituent \acp{ANN} of the interpolant was trained until an accuracy threshold was reached - with accuracy being measured as the percentage of the validation data set the \ac{ANN} was able to correctly predict (the meaning of ``correct'' is explained in Section~\ref{sec:method_interpolant_training}).

Details of the interpolant accuracy achieved during training are presented in Table~\ref{table:results_interpolant_accuracy}.

\begin{table}[ht!]
    \centering
    \caption{ANN training accuracy}
    \begin{tabular}{|E{4.3cm}|G{1.4cm}|G{1.4cm}|}
        \hline
        & \multicolumn{2}{c|}{\textbf{\#examples}} \\
        \cline{2-3}
        \multicolumn{1}{|c|}{} & \multicolumn{1}{c|}{\textbf{10}} & \multicolumn{1}{c|}{\textbf{100}} \\
        \hline
        \multicolumn{3}{|c|}{\textbf{\% correct}} \\
        \hline
        Highest & 100.0 & 100.0 \\
        \hline
        Lowest & 54.16 & 83.0 \\
        \hline
        Average & 99.44 & 99.58 \\
        \hline
        Stddev & 3.23 & 1.84 \\
        \hline
        \multicolumn{3}{|c|}{\textbf{Counts}} \\
        \hline
        \% correct $= 100$ & 139 & 97 \\
        \hline
        $99 \leq$ \% correct $< 100$ & 124 & 172 \\
        \hline
        $95 \leq$ \% correct $< 99$ & 26 & 20 \\
        \hline
        $90 \leq$ \% correct $< 95$ & 0 & 0 \\
        \hline
        $85 \leq$ \% correct $< 90$ & 0 & 1 \\
        \hline
        $80 \leq$ \% correct $< 85$ & 1 & 3 \\
        \hline
        \% correct $< 80$ & 2 & 0 \\
        \hline
        Number of networks evaluated & 292 & 293 \\
        \hline
    \end{tabular}
    \label{table:results_interpolant_accuracy}
\end{table}

The accuracy figures in Table~\ref{table:results_interpolant_accuracy} show that for both training datasets, the average accuracy achieved by the \acp{ANN} trained was better than 99.4\%, with more than 90\% of the \acp{ANN} achieving an accuracy of 99\% or better.

\begin{figure}
  \centering
  \includegraphics[width = 8.45cm]
    {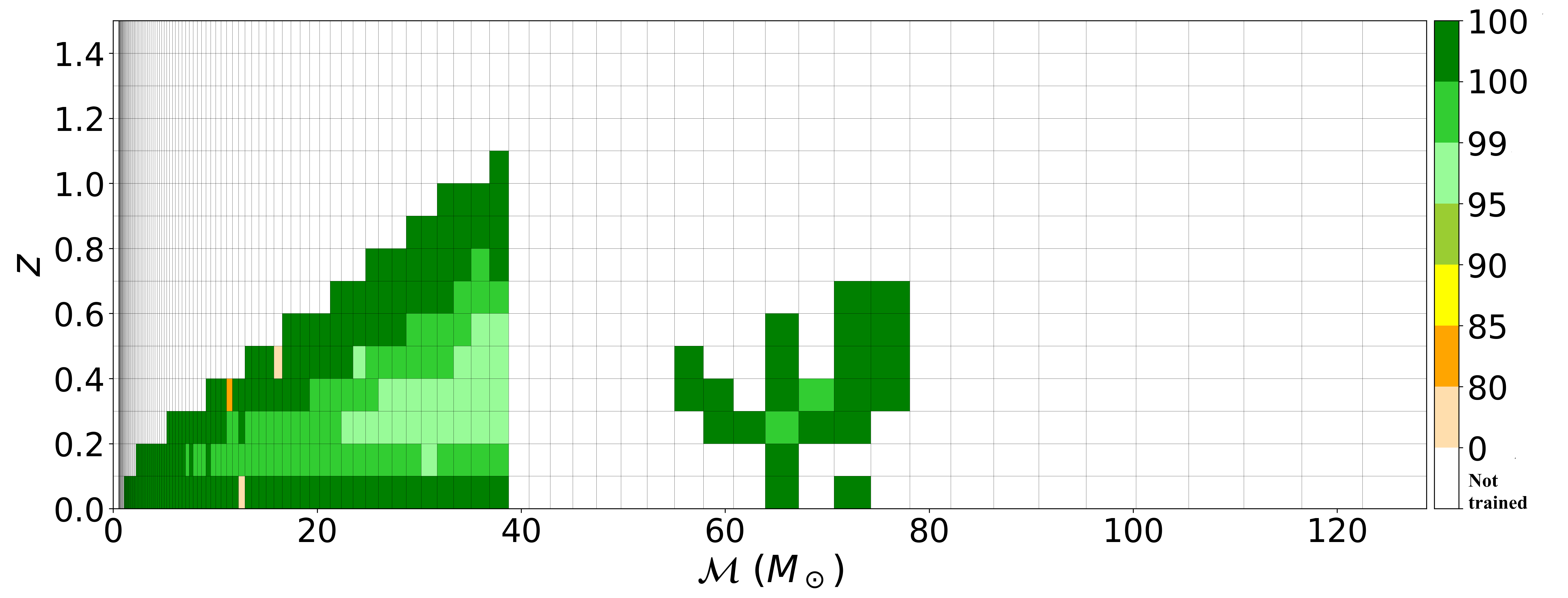}
    
  \includegraphics[width = 8.45cm]
    {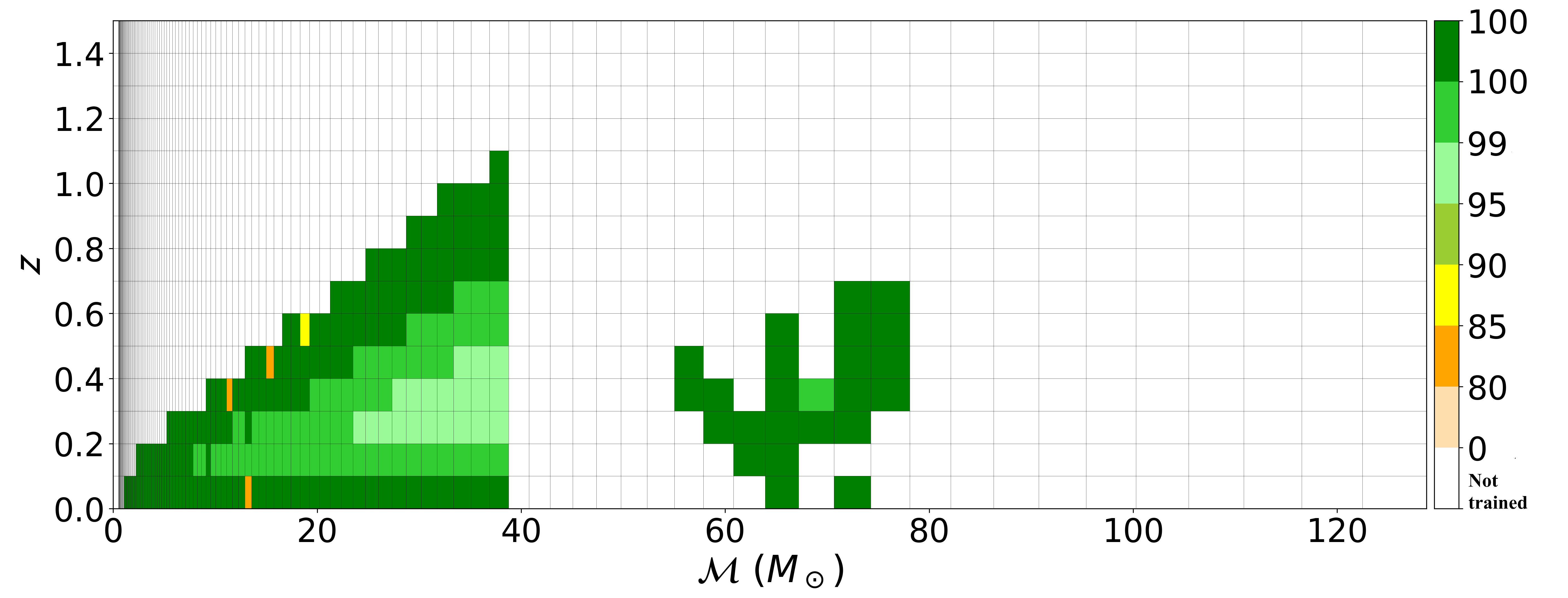}
    
      \includegraphics[width = 8.45cm]
    {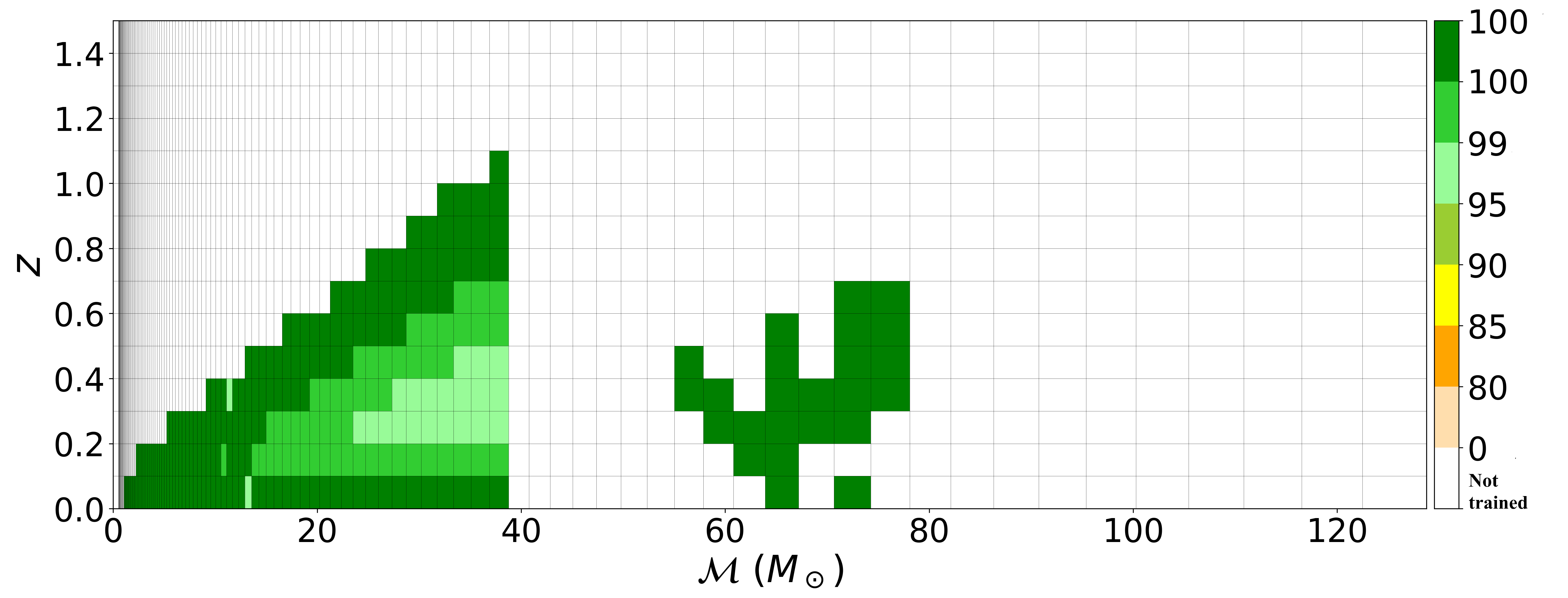}
    \caption{Interpolant training accuracy: 10 training examples (upper panel); 100 training examples \resp{(middle panel); 100 training examples with correctness threshold of $|t-p| \leq \max(\sqrt{t}/2, 0.5)$ (bottom panel)}.  White cells represent \acp{ANN} that did not require training (zero rates); the meaning of other colours is shown on the colour bar.}
    \label{fig:results_interpolant_accuracy}
\end{figure}

Figure~\ref{fig:results_interpolant_accuracy} shows in detail the accuracy achieved for each of the \acp{ANN} - the grid shown corresponds to the detection matrix, with each cell representing the \ac{ANN} corresponding to a \{$z, \mathcal{M}$\} bin.

From \resp{the top two panels of} Figure~\ref{fig:results_interpolant_accuracy} we can see that most of the \acp{ANN} (for both training datasets) were trained to a high degree of accuracy, with only a few \acp{ANN} not achieving 95\% accuracy or better. We can inspect the data and the training records for those few \acp{ANN}, and because the architecture of the interpolant allows us to easily replace individual constituent \acp{ANN} (or whatever model we have used for a particular cell), we can retrain them to achieve a better accuracy, change the \ac{ANN} architecture, etc.  \resp{For example, we find that increasing the number of hidden layers from 3 to 5 allows us to achieve 100\% accuracy across all cells.  However, there is a risk of attempting to memorise rather than learn the model with over-trained \acp{ANN} (see Appendix~\ref{sec:appendix_ANN_training_performance}).} 

The shape and location of the green areas in Figure~\ref{fig:results_interpolant_accuracy} is interesting. Recall that the white cells in Figure~\ref{fig:results_interpolant_accuracy} are \{$z, \mathcal{M}$\} bins for which the detection rate was below the threshold for zero (0.001 detections per year). The corollary is that the green areas are \{$z, \mathcal{M}$\} bins that contain \acp{DCO} that are detectable at LIGO O3 sensitivity. The triangle on the left is delineated by lower horizon distances for smaller masses along the diagonal line at the top, and by the lower edge of the pair-instability mass gap on the right (\citet{Heger_2002}, \citet{Marchant_2020}).  

\resp{Many of the bins along the top edge of the triangle (where the detection probability decreases toward zero at the horizon distance) and the bottom edge (only a small comoving volume is contained between redshifts of $z=0$ and $z=0.1$) have low detection rates.  These bins also contain the apparently poorly performing \acp{ANN} in the top two panels.  However, this is largely an artefact of our choice of a relative tolerance threshold for correctness, which becomes increasingly stringent as the detection rate decreases.  We can instead consider a combination of relative and absolute tolerances, which better accounts for the fact that a detection in a bin is unlikely if the predicted detection rate is low, making excessive precision in the surrogate model unnecessary.  If we replace the threshold for correctness of surrogate model prediction $p$ given expected target count detection count $t$ with $|t-p| \leq \max(\sqrt{t}/2, 0.5)$ (cf.~the default threshold of $|t-p| \leq \sqrt{t}/2$ from Section \ref{sec:method_interpolant_training}), we find that all \acp{ANN} achieve an accuracy above 95\% (see bottom panel of Figure \ref{fig:results_interpolant_accuracy}).}  

\begin{figure}
  \centering
  \includegraphics[width = 8.45cm]
    {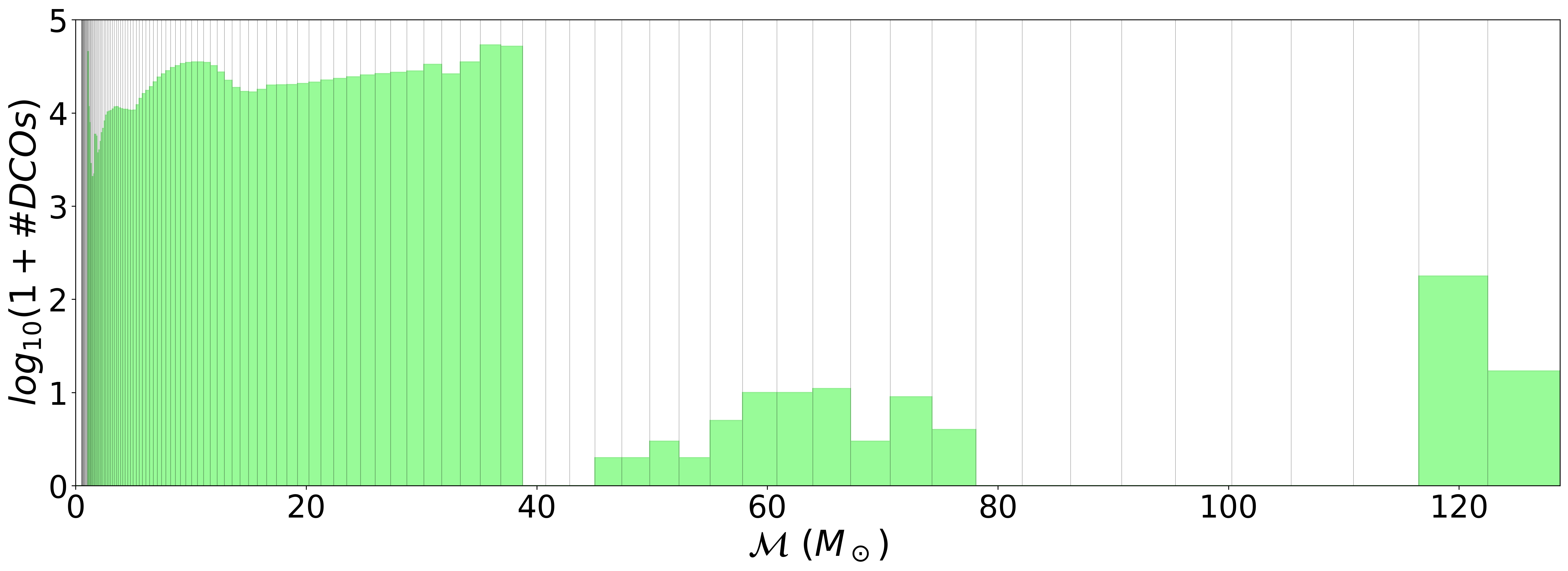}
    \caption{DCO counts for chirp mass $\mathcal{M}$ bins.}
    \label{fig:results_binCounts}
\end{figure}

We see a green cactus-shaped area to the right of the triangle (that is, to the right of the lower edge of the pair-instability mass gap). To understand this we need to understand the distribution of \acp{DCO} over the chirp mass range - this is shown in Figure~\ref{fig:results_binCounts}.  Figure~\ref{fig:results_binCounts} shows that almost all of the $\approx{1.3}$ million merging \acp{DCO} in the training data are located in the bins where $\mathcal{M} \lesssim {40}M_\odot$ (the green triangle in Figure~\ref{fig:results_interpolant_accuracy}), with a few bins in the $\approx{50}-80M_\odot$ range (the green cactus-shaped area) with 10 or fewer \acp{DCO} per bin, and then two bins at $\approx{120}M_\odot$ with a total of $\approx{190}$ \acp{DCO} (not seen in Figure~\ref{fig:results_interpolant_accuracy} - the detection rates for these bins were below the zero threshold).

Further inspection of the data reveals that of the $\approx{1.3}$ million merging \acp{DCO}, only 50 are located in the $\approx{50}-80M_\odot$ chirp mass range, and a further 193 are in the two chirp mass bins at $\approx{120}M_\odot$, and all of these binaries were BBHs comprised of two massive stars that began their evolution on the main sequence as chemically homogeneous stars \citep{Riley_2021}. Of the 50 binaries in the $\approx{50}-80M_\odot$ chirp mass range, 23 had both stars evolve as chemically homogeneous stars for the entirety of their lifetime on the main sequence, while for the remaining 27 the secondary star did not evolve as a chemically homogeneous star for its main sequence lifetime. Of the 193 binaries in the $\approx{120}M_\odot$ chirp mass range, all but one had both stars evolve as chemically homogeneous stars for the entirety of their lifetime on the main sequence - the secondary star of the remaining binary did not evolve as a chemically homogeneous star for its main sequence lifetime.

With only 243 of the $\approx{1.3}$ million \acp{DCO} produced from a population of 512 million binaries above the lower edge of the pair-instability mass gap, it is clear that these \acp{DCO} are very rare  (\citet{Marchant_2016}, \citet{Riley_2021}). The cactus shape in Figure~\ref{fig:results_interpolant_accuracy} is thus determined by the delay times  of chemically homogeneously evolving \acp{DCO} and convolution with the \ac{MSSFR}, with the detailed features likely an artefact of limited sampling.

\subsection{Inference with surrogate model}\label{sec:results_interpolant_performance}

\subsubsection{Method validation: inference on perfect measurements}\label{sec:results_method_validation}

In Section~\ref{sec:method_search} we outlined how we validate our method. Here we present the results of that validation.

Using \COMPAS post-processing tools and the synthesised population of 512 million binaries, we created a detection rate matrix, binned as described in Section~\ref{sec:method_binned_rates}, for a known value of $\lambda$: $\lambda(\alpha, \sigma, a_{SF}, d_{SF}) = (-0.325, 0.213, 0.012, 4.253)$.
Using that detection rate matrix and assuming an observing time of either 0.1 year or 1 year, we created mock data sets $\mathcal{D}$ with perfect parameter measurement accuracy for each of the ``detected'' sources (58 in the 0.1 year data set, 578 in the 1 year data set).

\begin{figure*}
  \centering
  \includegraphics[width = 17.5cm]{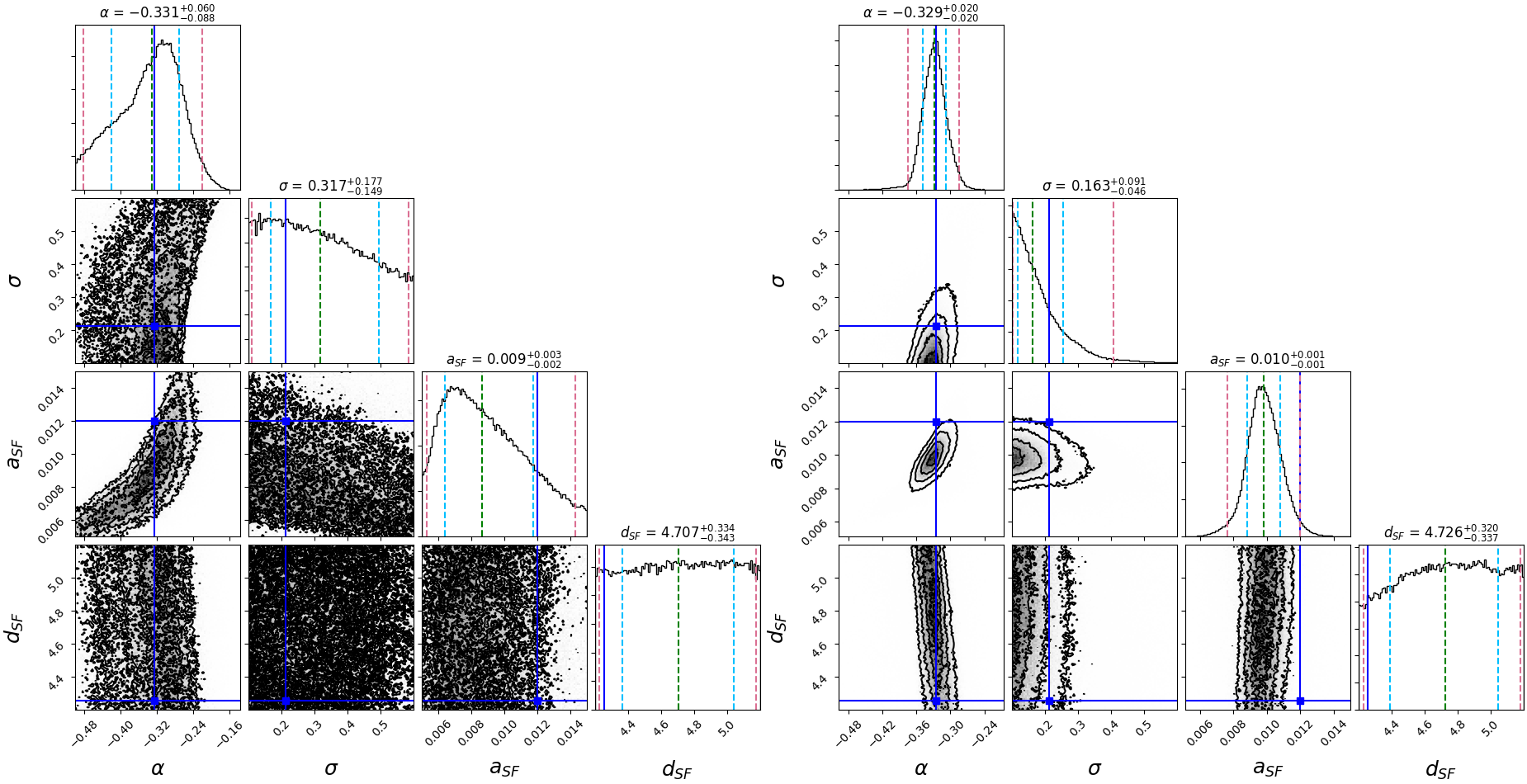}
    \caption{Corner plots showing \ac{MSSFR} parameter inference on mock observations without measurement uncertainties. The marginalised 1-d histograms for each \ac{MSSFR} parameter appear in the panels along the diagonal, and the marginalised 2-d projections of the posterior probability distributions in the off-diagonal panels. The left panel is for the smaller dataset (58 events); the right panel for the larger dataset (578 events). The blue solid line indicates the best-fit value, the green dashed line shows the median, and the pale-blue and pale-red dashed lines show the 68\% and 95\% credible intervals, respectively. Labels on the diagonals show the mean and 68\% credible interval.}
  \label{fig:results_mcmc_cornerplot_COMPAS}
\end{figure*}

We carried out a \ac{MCMC} search (e.g. \citet{Andrieu_2003}), using the \textit{emcee} Python package \citep{Foreman-Mackey_2013}, over the $\lambda=\{\alpha, \sigma, a_{SF}, d_{SF}\}$ parameter space, assuming flat priors on $\lambda$.  This search calculates the likelihood $\mathcal{L}\left(\mathcal{D}|\lambda_i\right)$ at each $\lambda_i$ visited by the \ac{MCMC} algorithm using the surrogate model (see Section~\ref{sec:method_likelihood_calculation} for the likelihood calculation). To confirm that our inference step was performing adequately, we also performed a fine-grained grid search consisting of 51 equidistant values for each of $\alpha$, $\sigma$, and $a_{SF}$, and 101 equidistant values for $d_{SF}$ - for a total of $51 \times 51 \times 51 \times 101 \approx 13.4$ million points evaluated, followed by a na\"ive hill-climbing search (e.g. \citet{Russell_Norvig:1995}), using the results of the grid search as a starting point.   We confirmed that the searches found the same maximum likelihoods (within expected sampling variations).

The \ac{MCMC} search posteriors are shown in Figure~\ref{fig:results_mcmc_cornerplot_COMPAS}. The true value is found within 68\% credible intervals for most \ac{MSSFR} parameters under study, and within 95\% credible intervals for all parameters.  As expected, the larger data set produces more precise inference on \ac{MSSFR} parameters, with the ratio of posterior width for $\alpha$ approaching a factor of $\sqrt{10}$ narrower on the data set with 10 times more data (right panel), though the improvement is much smaller for poorly measured parameters such as $d_{SF}$ whose posteriors are prior-dominated.

We considered surrogate models trained on data sets with 10 bootstrapped samples per $\lambda$ value  and 100 bootstrapped samples and found that they are not significantly different, which is consistent with the accuracy of the surrogate model shown in Figure~\ref{fig:results_interpolant_accuracy}.  We show results for surrogate models trained with 100 bootstrapped samples per $\lambda$ from here on.

\begin{figure*}
  \centering
  \includegraphics[width = 17.5cm]{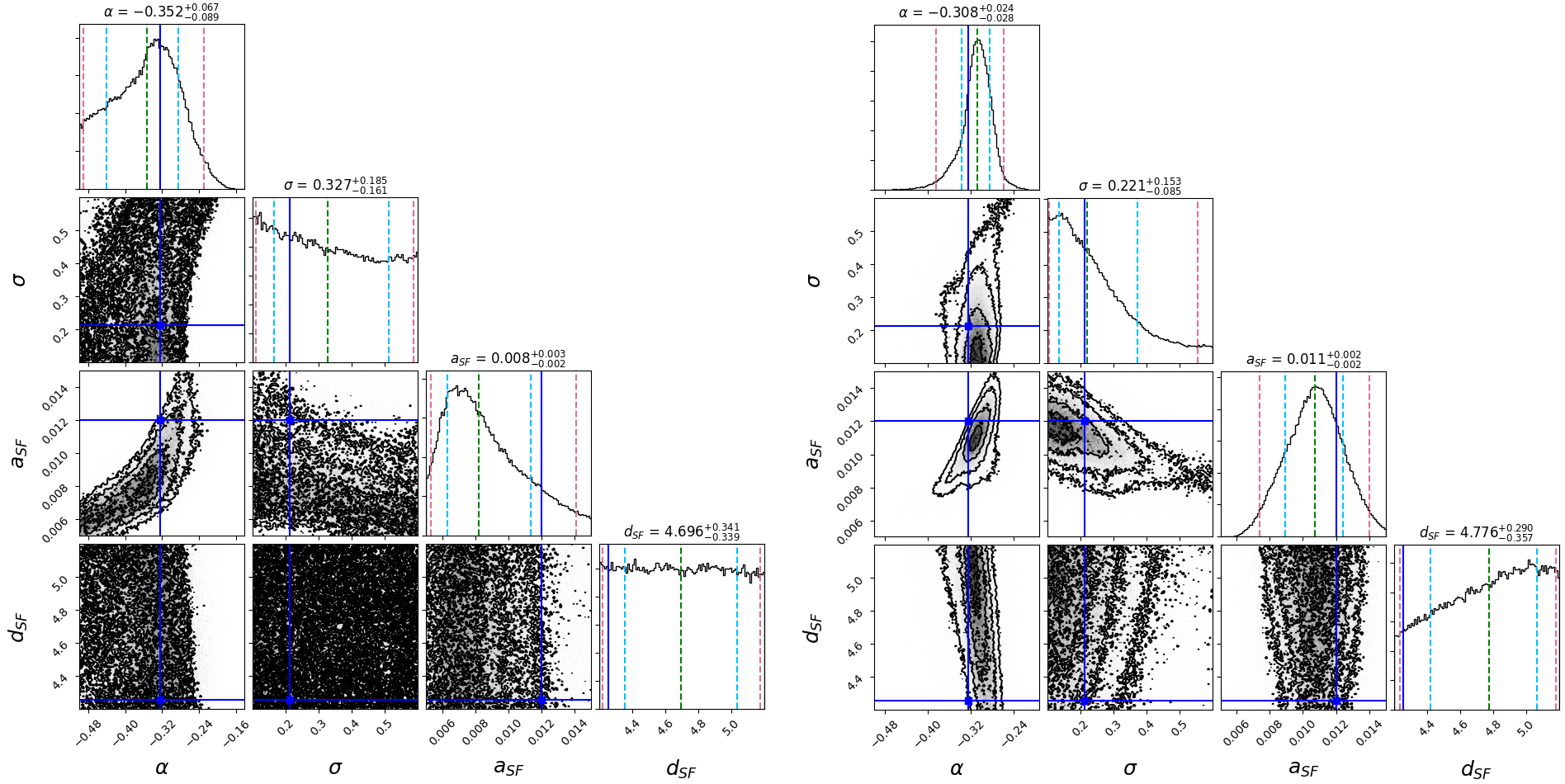}
    \caption{Corner plots for \ac{MSSFR} parameter inference on mock observations with measurement uncertainty. The left panel is for the smaller dataset (58 events); the right panel for the larger dataset (578 events). The blue solid line indicates the ``true'' value, the green dashed line shows the median, and the pale-blue and pale-red dashed lines show the 1$\sigma$ and 2$\sigma$ credible intervals respectively. }
  \label{fig:results_mcmc_cornerplot_mockLIGO}
\end{figure*}

\subsubsection{Method validation: inference on uncertain measurements}\label{sec:results_mock_LVK_inference}

We used the same detection rate matrix as in the previous subsection, corresponding to $\lambda(\alpha, \sigma, a_{SF}, d_{SF}) = (-0.325, 0.213, 0.012, 4.253)$, to validate inference on mock observations with measurement uncertainties. Using that detection rate matrix, assuming an observing time of one year, we created a dataset of mock LVK data containing 578 events. We then randomly sampled 58 events from that dataset, creating a new dataset with an observing time of 0.1 year.

We then replaced each of the chosen events with samples from an associated mock posterior.  To create these samples, we used a mock model of the LVK prior. We built the source-frame chirp mass prior by assuming that the component masses $m_1>m_2$ are uniformly drawn from the range $[1,1000] M_\odot$, with additional cuts $m_2 \in [0.05 m_1, m_1]$ and $\mathcal{M} \equiv m_1^{0.6} m_2^{0.6} (m_1+m_2)^{-0.2} \in [1,200] M_\odot$.  For the mock redshift prior, we used $\pi(z) \propto z^{2}$ on $z \in [0.01,1.5]$. We then weighed mock chirp mass and redshift samples taken from 

\begin{eqnarray}\label{eq:mockposterior}
\mathcal{M}  \sim \mathcal{M}^T (1+0.03\frac{12}{\rho} (r_0 +r));\\
z  \sim z^T (1+0.3\frac{12}{\rho} (r_0 +r)) \nonumber
\end{eqnarray}

by these priors. In Eq.~\ref{eq:mockposterior}, which follows \citet{Powell_2019}, the superscript $^T$ denotes the true value, $\rho$ is the signal-to-noise ratio sampled from $p(\rho) \propto \rho^{-4}$ with a minimum of $\rho \geq 12$, $r_0$ is a normal random variable which stochastically shifts the peak of the posterior away from the truth, and $r$ is a vector of normal random variables which provides the spread of the posterior.  Thus, the small and large data sets are composed of 58 and 578 events, each represented by a set of mock posterior samples that contain mock measurement uncertainties on the observed parameters $\mathcal{M}$ and $z$.

Using each of those datasets as ``true'' data, we used our surrogate model to infer posteriors on $\lambda$ assuming flat priors. The \ac{MCMC} search statistics are shown in Figure~\ref{fig:results_mcmc_cornerplot_mockLIGO}.

Figure~\ref{fig:results_mcmc_cornerplot_mockLIGO} shows that, as expected, inference provides consistent credible intervals, with most ``true'' values falling within the 1-$\sigma$ credible interval, and the remainder within the 2-$\sigma$ credible interval. As before, the larger data set produces more precise inference on \ac{MSSFR} parameters.  Comparing Figure~\ref{fig:results_mcmc_cornerplot_mockLIGO} with Figure~\ref{fig:results_mcmc_cornerplot_COMPAS}, we see that mock measurement uncertainties have limited impact on \ac{MSSFR} inference for the smaller data set, where inference is predominantly limited by the total number of events.  For the larger data set, incorporating measurement uncertainty does lead to a moderate deterioration in the accuracy of inference.  

\begin{figure*}
  \centering
  \includegraphics[width = 17.5cm] 
  {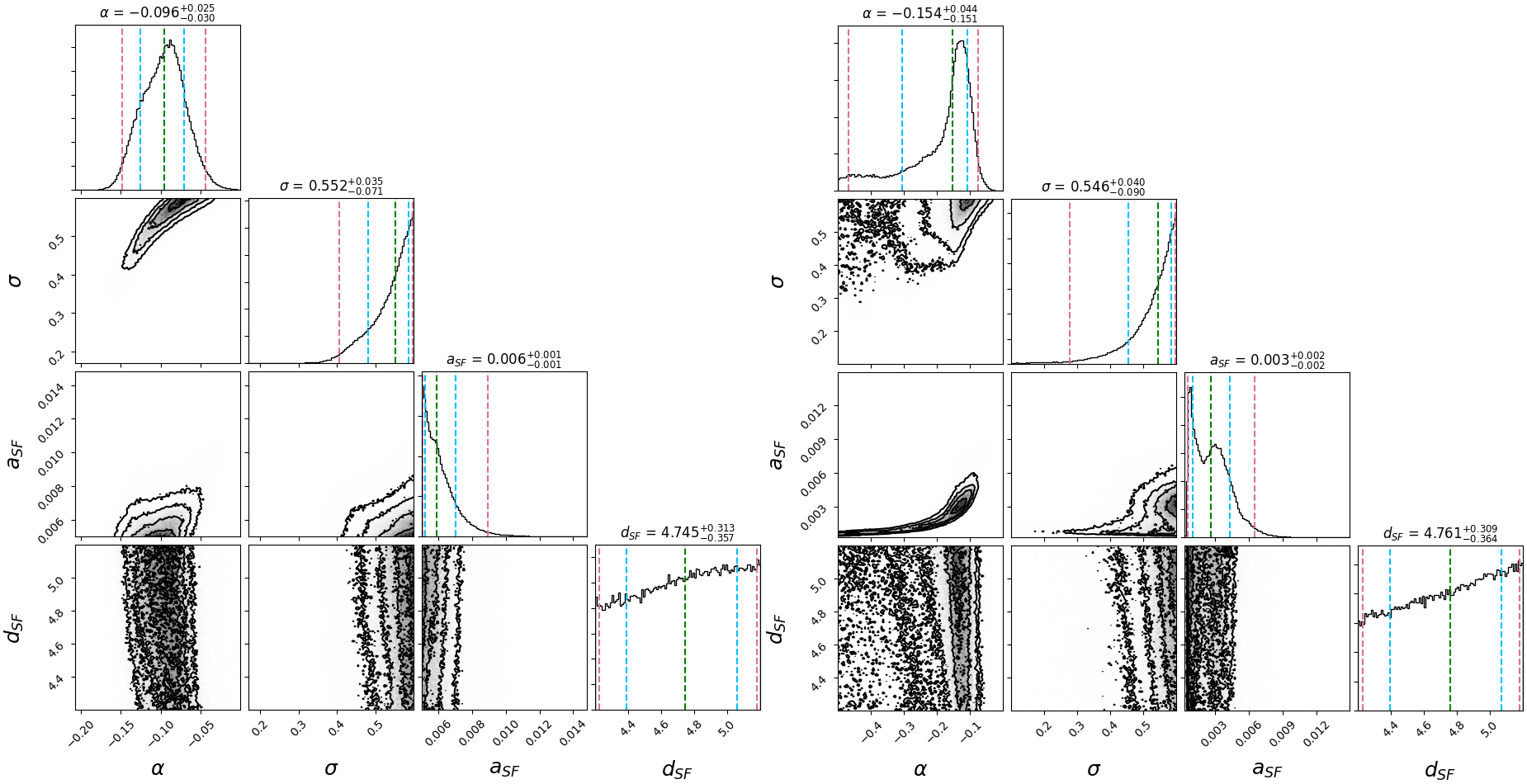}
    \caption{Corner plot for \ac{MSSFR} parameter inference on LVK data with $p_{astro} \ge 0.95$ using the original priors (left panel) and using the extended prior on $a_{SF}$ (right panel).}
  \label{fig:results_mcmc_cornerplot_LIGO}
\end{figure*}

\begin{figure*}
  \centering
  \includegraphics[width = 17.5cm]{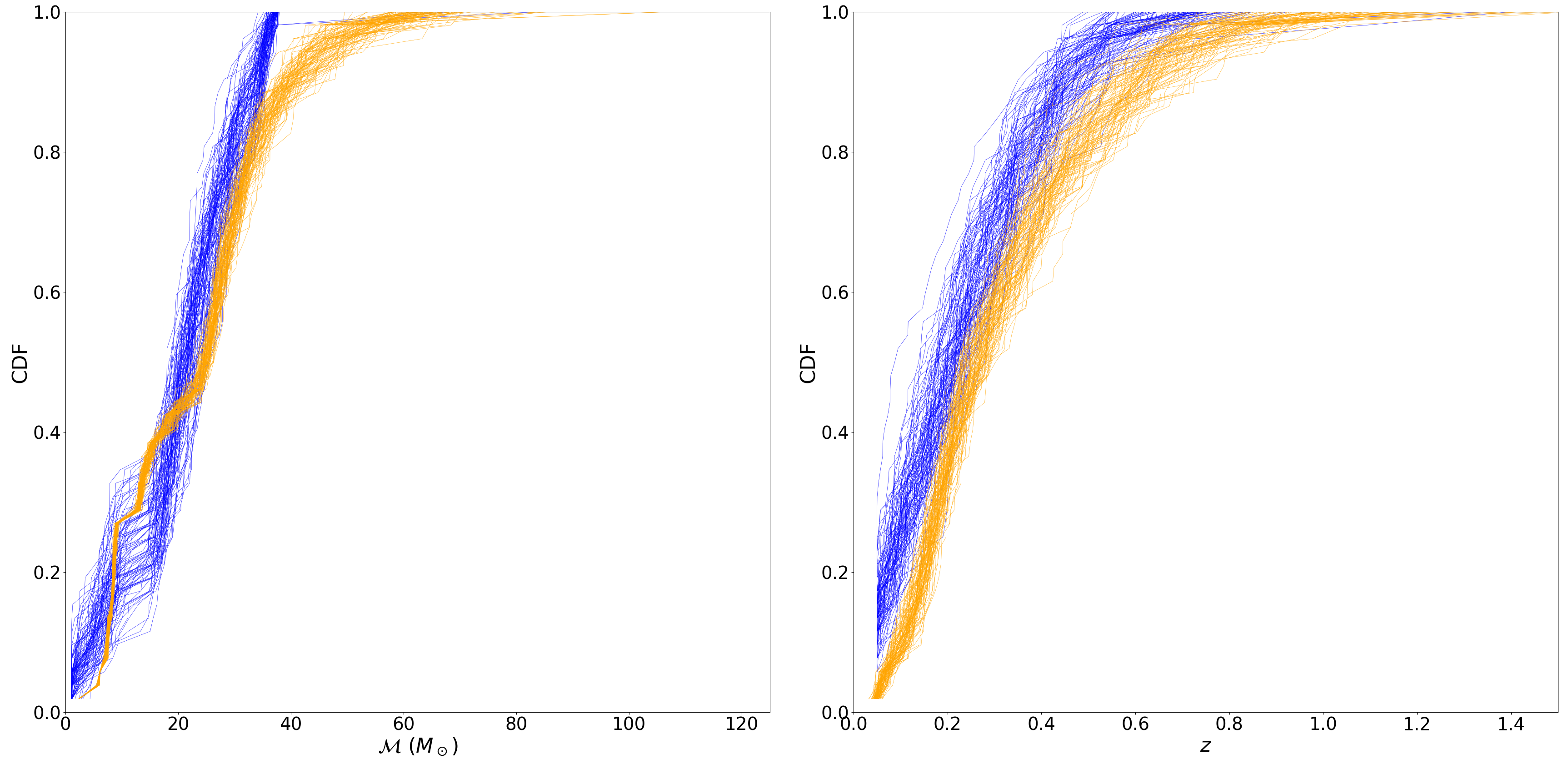}
    \caption{Cumulative Density Functions for chirp mass (left panel) and redshift (right panel).
    In each panel the \resp{100 orange curves are} randomly sampled cumulative density functions from the posterior samples (with LVK priors) for each of the \ac{DCO} mergers detected during the third observing run ($p_{astro} \ge 0.95$), and indicates the spread due to measurement uncertainty. The  \resp{100} blue curves are cumulative density functions constructed by sampling 52 events from the predicted distribution of detectable \ac{DCO} mergers from the best-fit $\lambda$ and indicate the sample variance.}
  \label{fig:results_CDFs_LIGO}
\end{figure*}

\subsubsection{Inference from real LVK observations}\label{sec:results_real_LVK_inference}

Since the goal of this work is to (eventually) develop a method to allow us to infer constraints for (some of) the astrophysical states and processes, as an illustration of the ability of this methodology to assist in inferring constraints on cosmological and astrophysical parameters from real data, we performed an \ac{MCMC} search (as described in Section~\ref{sec:results_method_validation}) on actual LVK observations from observing runs 3a and 3b (run duration 275.3 days), where $p_{astro} \ge 0.95$ (52 events).

Here, we present the results of this analysis.  However, a few disclaimers are warranted.  We only varied parameters in a \ac{MSSFR} model, while fixing the assumptions and parameters describing stellar and binary evolution (mass transfer, winds, supernovae, etc.) to fiducial, but likely incorrect, values.  We also used a simple approximation for the selection function imposed by LVK observations, picking a sample noise spectrum rather than a variable one and using a single-detector signal-to-noise ratio threshold of 8, independent of the source parameters, as a proxy for detectability by the network.  A more realistic selection function would be based on an actual injection campaign into the detector noise as accumulated over the third LVK observing run.

The posteriors on the \ac{MSSFR} parameters on the LVK O3 data are shown in the left panel of Figure~\ref{fig:results_mcmc_cornerplot_LIGO}.  The highest-likelihood point found by the \ac{MCMC} chain is at $\lambda(\alpha, \sigma, a_{SF}, d_{SF}) = (-0.085, 0.598, 0.005, 5.042)$.  It is clear that the posteriors rail against the prior boundary on $a_{SF}$.

We therefore repeated the analysis by extending the uniform prior on $a_{SF}$ to incorporate the range $[0,0.015]$.  Fortunately, $a_{SF}$ is a scaling factor only, so this did not require re-training the neural network.  We used the surrogate model prediction at the mid-point of the $a_{SF}$ range over which it was trained to predict the detection rate matrix at values of $a_{SF}$ outside this range.  Specifically, we assumed that the surrogate model for the detection rate matrix at $a_{SF}<0.005$ was equal to $(a_{SF} / 0.01)$ times the prediction of the surrogate model at $\{a_{SF}=0.01, \lambda'\}$, where $\lambda'\equiv\lambda/a_{SF}$ refer to all other parameters.  We show the posteriors on \ac{MSSFR} parameters with this extended prior in the right panel of Figure~\ref{fig:results_mcmc_cornerplot_LIGO}.  The highest-likelihood point found by the \ac{MCMC} chain with extended priors is at $\lambda(\alpha, \sigma, a_{SF}, d_{SF}) = (-0.115, 0.599, 0.003, 5.158).$ These posteriors rail against the prior on $\sigma$, suggesting that an even broader distribution of metallicities is preferred under this \ac{MSSFR} model.  On the other hand, the preferred value of the normalisation of the star-formation rate $a_{SF}$ is a factor of $\sim 5$ lower than the estimate of \citet{Madau_2014}.

\begin{figure*}
  \centering
  \includegraphics[width = 17.5cm]{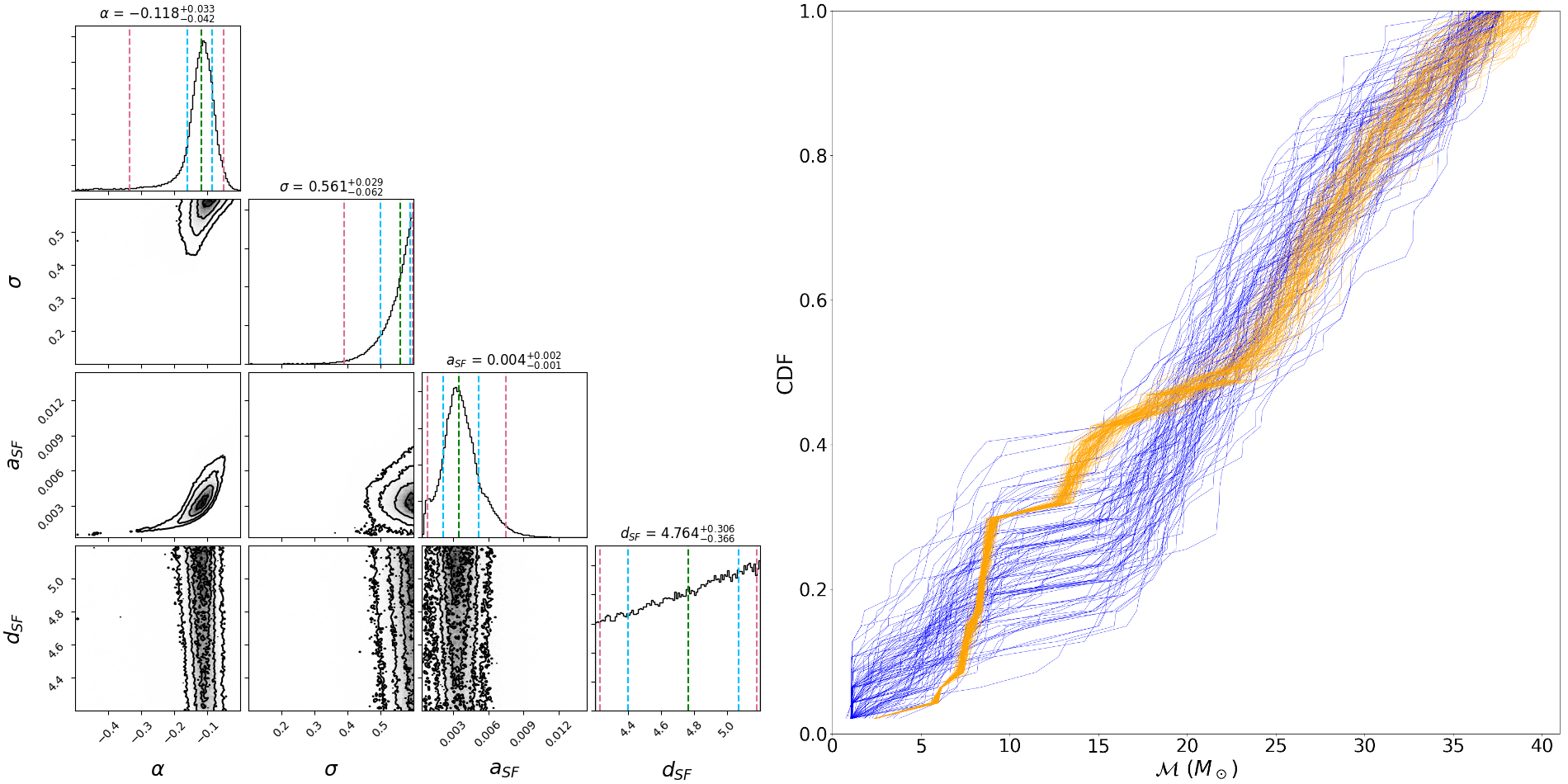}
    \caption{\resp{Inference on LVK data after removing the 5 most massive binaries from the data set (see section \ref{sec:results_real_LVK_inference}).  Left panel: Corner plot for \ac{MSSFR} parameter inference on LVK data, as in Figure \ref{fig:results_mcmc_cornerplot_LIGO}; right panel: posterior predictive check on the chirp mass distribution for the associated best-fit parameters, as in Figure \ref{fig:results_CDFs_LIGO}.}}
  \label{fig:results_corner_CDFs_LIGO_40msun}
\end{figure*}

In Figure~\ref{fig:results_CDFs_LIGO} we show the cumulative density functions (CDFs) for chirp mass (left panel) and redshift (right panel), for both LVK observations and for our best-fit $\lambda$ (from the analysis with the extended prior on $a_{SF}$).  This provides a visual posterior predictive check. In this case, our preferred \ac{MSSFR} model fails this check.  The paucity of higher-redshift observations in the model prediction likely indicates an overly conservative detectability estimate, probably through the use of a pessimistic threshold on detectable signal-to-noise ratio and/or a pessimistic noise power spectral density estimate.  While this problem is relatively easy to fix, the failure to reproduce the observed features of the chirp mass distribution, such as the sharp, narrow peak near $\mathcal{M} \approx 10\ M_\odot$, speaks to a flawed underlying binary evolution model.  As noted above, we did not consider variations in the stellar and binary evolution assumptions and model parameters, as this was only intended as a proof-of-concept exercise.  

\resp{Moreover, some of the events, particularly the most massive ones, have been conjectured to arise from channels other than isolation binary evolution, which we did not consider here (see, e.g., \citealt{MandelFarmer_2018} for a review).  Other channels may be particularly relevant for higher-mass binaries, including hierarchical dynamical mergers \citep[e.g.,][]{GerosaFishbach_2021} or mergers in AGN disks in which black holes could grow by accretion \citep[e.g.,][]{Tagawa_2021}.  We therefore consider an alternative data set in which we remove the five LVK binaries whose median source-frame chirp mass exceed $40 M_\odot$ and reweigh the posteriors for remaining binaries by imposing a strict chirp mass prior upper limit of $40 M_\odot$.  We also remove any binaries with chirp mass exceeding $40 M_\odot$ from the forward models.  The resulting posteriors on \ac{MSSFR} parameters (using the extended prior on $a_{SF}$) and the cumulative posterior predictive check on the chirp mass distribution are shown in Figure \ref{fig:results_corner_CDFs_LIGO_40msun}.  The posteriors on \ac{MSSFR} parameters shift within one standard deviation relative to Figure \ref{fig:results_mcmc_cornerplot_LIGO}, allowing for a more consistent chirp mass posterior predictive check than in Figure~\ref{fig:results_CDFs_LIGO}, although our model still fails to reproduce the narrow peak near $\mathcal{M} \approx 10\ M_\odot$.  However, this analysis is based on a single model of binary evolution and only considers the impact of \ac{MSSFR} parameters; complete inference must simultaneously incorporate the parameters governing stellar and binary evolution along with \ac{MSSFR} models.}

\subsubsection{Inference performance}\label{sec:results_inference_performance}

All \ac{MCMC} searches performed for this study used an ensemble of 10 walkers (single-threaded), and were allowed to run for 100,000 steps per walker, resulting in 1,000,000 steps per search.  Each step involves the creation of a new detection rate matrix (for the values of $\lambda$ visited at that step). We used the \acp{ANN} trained on the 100-sample dataset, so our surrogate model was comprised of 293 \acp{ANN}, with an execution time of ${\sim}6.42 \times 10^{-3}$ seconds (Section~\ref{sec:method_interpolant_performance}).

The elapsed time for each \ac{MCMC} search using our surrogate model is ${\sim}6.42 \times 10^{-3} \times 1,000,000 = {\sim}{6,420}$ seconds (${\sim}1.8$ hours), ignoring \ac{MCMC} algorithm overheads, which are small compared to the execution time of the \acp{ANN}. The time to create all training data sets was 142 hours, and the time to train the \acp{ANN} was 12 hours, for a total time 156 hours.

The \COMPAS post-processing tools execution time to create a single detection rate matrix is ${\sim}112$ seconds (Section~\ref{sec:method_interpolant_performance}). The elapsed time for equivalent \ac{MCMC} searches using the \COMPAS post-processing tools to create the detection rate matrices is ${\sim}112 \times 1,000,000 = {\sim}{112,000,000}$ seconds (${\sim}3.5$ years), ignoring \ac{MCMC} algorithm overheads, which are small compared to the execution time of the \COMPAS post-processing tools. Thus, the use of surrogate models reduced the total inference cost by a factor of 200 after accounting for the time needed to create training data and train these models.

%% file: conclusion.tex
\section{Concluding remarks}\label{sec:conclusion}

We conducted a proof-of-concept study that showed that it is possible to construct an interpolant that can calculate \ac{DCO} merger detection rates for a set of astrophysical parameters, with fairly good accuracy, and in milliseconds (several orders of magnitude faster than methods currently available). This allows us to create a large synthetic state-space in a very reasonable timeframe, which can then be used to conduct inference about the parameters. 

The method we developed is fast, flexible, highly parallelisable, and robust - the constituent \acp{ANN} can be retrained, or replaced, individually and as necessary. We achieved a total analysis cost reduction by a factor of 200 after accounting for surrogate model training.

We found that our first attempt at analysis on LVK data pushed our inference to rail against the prior boundaries.  In this case this was likely due to poor choices in the fiducial model describing astrophysical evolution, exacerbated by an overly simplistic observational selection function.   We were able to partially remedy this issue because one of the parameters required only a simple rescaling.  However, to avoid similar problems in future analyses, it would be wise to use active learning and grow the training data set in regions of parameter space which provide models with the best match to the data.

For this study we focussed on just the four \ac{MSSFR} parameters used in the post-processing of \COMPAS simulations, but we are not limited to those parameters, or to the post-processing code. The next step is to expand this work to other astrophysical parameters that govern stellar and binary evolution in the population synthesis code itself (e.g. \COMPAS), rather than just the post-processing code, thereby reducing the need for time-consuming simulations.

%% file: ack.tex
\section*{Acknowledgements}\label{sec:ack}
Simulations in this paper made use of the \COMPAS rapid population synthesis code, which is freely available at \url{http://github.com/TeamCOMPAS/COMPAS}.  
The version of \COMPAS used for these simulations was v02.31.00. 

IM is a recipient of the Australian Research Council Future Fellowship FT190100574.

%% file: appendix-ANNs.tex
\section{ANN training and performance}\label{sec:appendix_ANN_training_performance}

We did not use extremely large datasets to train the \acp{ANN} that comprise the interpolant in this work. Even so, we achieved very good results with the \acp{ANN} we trained - most achieved 100\% accuracy when tested on the training and validation data. These results might lead us to wonder whether the \acp{ANN} have been over-fitted during training (though we did test that they are capable of generalising).

How many training examples are required to properly train an \ac{ANN}? The answer depends on the complexity of the problem, and the \ac{ANN} being used to solve it. It is effectively unknowable \textit{a priori}, and must be discovered through empirical investigation. In other words, we can tell whether the training data were sufficient by testing that the \ac{ANN} gets the right answer for the examples it was trained on, and whether it predict correct answers for inputs it was not trained / validated on.  But were the data necessary?

The conventional wisdom is that at least thousands of training examples are required, certainly no fewer than hundreds. Tens, or hundreds, of thousands for an "average" modelling problem; millions, or tens of millions, for a "hard" problem. This basically translates to  "Get as much data as possible and use it" - not an unreasonable strategy in a situation where a better answer is not immediately apparent.

We know that when a sufficiently large \ac{ANN} tries to learn from a small dataset it will tend to memorise the entire dataset - an example of over-fitting. But does this compromise the ability of the network to generalise beyond the the training dataset? In the past we might have thought so, but recent work suggests that this is not necessarily the case (\citet{Zhang_2017}, \citet{Zhang_2021}).

\citet{Mukherjee_2022} claim that the properties of a particular mathematical operation, \textit{expand-and-sparsify} \citep{Dasgupta_2020}, explain the ability of an over-parameterised \ac{ANN} (i.e. the number of parameters in the model exceeds the size of the training dataset) to both memorise the entire training dataset \textit{and} generalise beyond the training dataset (by learning the underlying structure to the training data).

As noted above, we achieved very good results with the \acp{ANN} we trained, yet in most cases the number of epochs over which the networks were trained, and so the time to train, was not particularly onerous. This, especially when coupled with the performance achieved by the \acp{ANN} when trained on a relatively small dataset, raises the question of why it wasn't more difficult and time-consuming to find a solution to such a difficult problem. The answer is two-fold.

First, the phenomenon of untrained networks with randomly-initialised weights performing surprisingly well has been reported in the past (e.g. \citet{Zhang_2022}, \citet{Frankle_2020}). This is often explained by referring to the Lottery Ticket Hypothesis \citep{Frankle_2018}, which posits that \acp{ANN} are really running large lotteries where sub-networks whose weights have been serendipitously randomly-initialised to values that produce good results are scattered throughout the network. Thus, in a sufficiently large, untrained, and randomly-initialised \ac{ANN}, a sub-network with random weights that performs as well as a trained network can be found, and the remainder of the network is ignored.

Next, as noted in Section~\ref{sec:method_interpolant_training}, we trained the \acp{ANN} using the Keras Adam optimiser, which uses an enhanced stochastic gradient descent algorithm. We know that the standard stochastic gradient descent algorithm is usually sufficient to train an \ac{ANN} - it is not often that other methods of optimisation for the network parameters are required. \citet{Chizat_2019} note that often during training, the parameters of an \ac{ANN} change very little from their initialised values. Mukherjee and Huberman claim that since very often the randomly-initialised weights are "good enough" (as described above), they require hardly any fine-tuning for better performance, so an optimiser such as stochastic gradient descent needs to do very little work to train a network to achieve reasonably good performance. Indeed, \citet{Mukherjee_2022} contend that the training of an \ac{ANN} is the third most important component of a successful \ac{ANN} solution, behind architecture selection and parameter initialisation.

%% file: appendix-COMPAS_fiducial.tex
\section{COMPAS configuration fiducial values}\label{sec:appendix_COMPAS_fiducial}

\begin{table*}[ht!]
\centering
\caption{Initial values and default settings for the \COMPAS fiducial model.}
\label{tab:app_COMPAS_fiducial}
\resizebox{\textwidth}{!}{%
\begin{tabular}{lll}
\hline  \hline
Description and name                                 														& Value/range                       & Note / setting   \\ \hline  \hline
\multicolumn{3}{c}{Initial conditions}                                                                      \\ \hline
Initial primary mass \monei                               															& $[5, 150]$\Msun    & \citet{Kroupa_2001} IMF  $\propto  {\monei}^{-\alpha}$  with $\alpha_{\rm{IMF}} = 2.3$ for stars above $5$\Msun	  \\
Initial mass ratio $\qi = \mtwoi / \monei $           												& $[0, 1]$                          &       We assume a flat mass ratio distribution  $p(\qi) \propto  1$ with \mtwoi $\geq 0.1\Msun$   \\
Initial semi-major axis \ai                                            											& $[0.01, 1000]$\AU & Distributed flat-in-log $p(\ai) \propto 1 / {\ai}$ \\   
Initial metallicity \Zi                                           											& $[0.0001, 0.03]$                 & Distributed uniform-in-log   \\
Initial orbital eccentricity \ei                                 							 				& 0                                & All binaries are assumed to be circular at birth  \\
%
\hline
\multicolumn{3}{c}{Fiducial parameter settings:}                                                            \\ \hline
Stellar winds  for hydrogen rich stars                                   																&      \citet{Belczynski_2010a}    &   Based on {\citet{Vink_2000, Vink_2001}}, including  LBV wind mass loss with $f_{\rm{LBV}} = 1.5$   \\
Stellar winds for hydrogen-poor helium stars &  \citet{Belczynski_2010a} & Based on   {\citet{Hamann_1998}} and  {\citealt{Vink_2005}}  \\

%
Max transfer stability criteria & $\zeta$-prescription & Based on \citet[][]{Vigna-Gomez_2018} and references therein     \\ 
Mass transfer accretion rate & thermal timescale & Limited by thermal timescale for stars  \citet[][]{Hurley_2002, Vinciguerra_2020} \\ 
 & Eddington-limited  & Accretion rate is Eddington-limit for compact objects  \\
Non-conservative mass loss & isotropic re-emission &  {\citet[][]{Massevitch_1975, Bhattacharya_1991, Soberman_1997}} \\ 
& &  {\citet{Tauris_2006}} \\
Case BB mass transfer stability                                														& always stable         &       Based on  \citet{Tauris_2015, Tauris_2017, Vigna-Gomez_2018}         \\ 
%
%
CE prescription & $\alpha-\lambda$ & Based on  \citet{Webbink_1984, deKool_1990}  \\
CE efficiency $\alpha$-parameter                     												& 1.0                               &              \\
CE $\lambda$-parameter                               													& $\lambda_{\rm{Nanjing}}$                             &        Based on \citet{Xu_2010a, Xu_2010b} and  \citet{Dominik_2012}       \\
Hertzsprung gap (HG) donor in {CE}                       														& pessimistic                       &  Defined in \citet{Dominik_2012}:  HG donors don't survive a {CE}  phase        \\
%
%
{SN} natal kick magnitude \vk                          									& $[0, \infty)$\kms & Drawn from Maxwellian distribution with standard deviation $\sigma_{\rm{rms}}^{\rm{1D}}$          \\
 {SN} natal kick polar angle $\thetak$          											& $[0, \pi]$                        & $p(\thetak) = \sin(\thetak)/2$ \\
 {SN} natal kick azimuthal angle $\phi_k$                           					  	& $[0, 2\pi]$                        & Uniform $p(\phi) = 1/ (2 \pi)$   \\
 {SN} mean anomaly of the orbit                    											&     $[0, 2\pi]$                             & Uniformly distributed  \\
Core-collapse  {SN} remnant mass prescription          									     &  delayed                     &  From \citet{Fryer_2012}, which  has no lower {BH} mass gap  \\%
USSN  remnant mass prescription          									     &  delayed                     &  From \citet{Fryer_2012}   \\%
ECSN  remnant mass presciption                        												&                                 $m_{\rm{f}} = 1.26\Msun$ &      Based on Equation~8 in \citet{Timmes_1996}          \\
Core-collapse  {SN}  velocity dispersion $\sigma_{\rm{rms}}^{\rm{1D}}$ 			& 265\kms           & 1D rms value based on              \citet{Hobbs_2005}                          \\
 USSN  and ECSN  velocity dispersion $\sigma_{\rm{rms}}^{\rm{1D}}$ 							 	& 30\kms             &            1D rms value based on e.g.,    \citet{Pfahl_2002, Podsiadlowski_2004}    \\
PISN / PPISN remnant mass prescription               											& \citet{Marchant_2019}                    &       As implemented in \citet{Stevenson_2019}      \\
Maximum NS mass                                      & $\rm{max}_{\rm{NS}} = 2.5$\Msun & Following \citet{Fryer_2012}            \\
Tides and rotation & & We do not include tides and/or rotation in this study\\
Binary fraction                                      & $f_{\rm{bin}} = 0.7$ &  \\
Solar metallicity \Zsun                             & $\rm{Z}_{\odot}\xspace = 0.0142$ & based on {\citet{Asplund_2009}} \\
\hline
\multicolumn{3}{c}{Simulation settings}                                                                     \\ \hline
Binary population synthesis code                                      & COMPAS &       \citet{Stevenson_2017, Barrett_2018, Vigna-Gomez_2018, Neijssel_2019} \\
& & \citet{Broekgaarden_2019, COMPAS_2022}.        \\
\hline \hline
\end{tabular}%
}
\end{table*}